\documentclass[preprint,authoryear,12pt]{elsarticle_preprint}

\usepackage[svgnames]{xcolor}
\usepackage{graphicx, multirow, subfigure}
\usepackage{amsmath, amssymb, amsthm}
\journal{Computational Statistics \& Data Analysis}

\newtheorem{theorem}{Theorem}[section]

\theoremstyle{remark}
\newtheorem{example}{Example}[section]

\begin{document}
\begin{frontmatter}

\title{Spline-backfitted kernel forecasting for \\functional-coefficient autoregressive models}
\author{Joshua D.~Patrick\corref{cor1}}
\address{Department of Statistics,
One Shields Avenue, University of California, 
Davis, CA 95616-5270}
\ead{jdpatrick@ucdavis.edu}
\cortext[cor1]{Corresponding author}
\author{Jane L.~Harvill}
\address{Department of Statistical Science,
P.O.~Box 97140,
Baylor University,
Waco, TX 76798-7140}
\ead{jane\_harvill@baylor.edu}
\author{Justin R.~Sims}
\address{Department of Statistical Science,
P.O.~Box 97140,
Baylor University,
Waco, TX 76798-7140}
\ead{justin\_sims@baylor.edu}

\begin{abstract}
We propose three methods for forecasting a time series modeled using a functional coefficient autoregressive model (FCAR) fit via spline-backfitted local linear (SBLL) smoothing.  The three methods are a ``naive" plug-in method, a bootstrap method, and a multistage method. We present asymptotic results of the SBLL estimation method for FCAR models and show the estimators are oracally efficient. The three forecasting methods are compared through simulation. We find that the naive method performs just as well as the multistage method and even outperforms it in some situations. We apply the naive and multistage methods to solar irradiance data and compare forecasts based on our method to those of a linear AR model, the model most commonly applied in the solar energy literature.  
\end{abstract}

\end{frontmatter}

\section{Introduction}

There are many useful time series models that lie between the class of linear, fully parametric models, and nonlinear nonparametric models.  One such model is the functional coefficient autoregressive (FCAR) model, defined as
\begin{equation}
X_{t}=\sum_{\alpha=1}^{p}m_{\alpha}\left(U_{t-d}\right)X_{t-\alpha}+\sigma\left(\textbf{V}_t,\textbf{X}_{t}\right)\varepsilon_{t},\qquad t=1,\ldots,n,
\label{eq:fcar1}
\end{equation}
where $p$ is a positive integer, $m_{\alpha}\left(U_t\right)$
is a measurable function of $U_t$,  for $\alpha=1,\ldots,p$,
$\sigma^{2}\left(\textbf{V}_t,\textbf{X}_{t}\right)$ is a variance function dependent on $\textbf{V}_{t}=\left(X_{1},\ldots,X_{n-d}\right)^{\prime}$ and $\textbf{X}_{t}=\left(X_{1},\ldots,X_{n}\right)^{\prime}$, and $\left\{ \varepsilon_{t}\right\} $ is a sequence of i.i.d.~random
variables with mean $0$ and variance $\sigma^2 < \infty$.  Usually the variable $U_t$ is taken to be a lagged value of the series; i.e., $U_t = X_{t-d}$, where $d$ is a positive integer.  Although the FCAR model imposes an autoregressive structure, its flexibility lies in allowing the autoregressive coefficients $m_\alpha$, to vary as a function of $U_t$.  While reducing the size of the class of nonlinear models, the class of FCAR models is broad enough to include some common nonlinear time series models as specific cases.  Among these are the threshold autoregressive (TAR) model of \cite{tong1983}, the exponential autoregressive (EXPAR) model of \cite{haggan1981}, and the smooth transition autoregressive (STAR) model of  \cite{chan1986}.

\cite{chen1993} introduced the FCAR model and proposed a procedure for building the model based on arranged local autoregression which constructs estimators based on an iterative recursive formula that resembles local constant smoothing. They compared the proposed model building procedure to threshold and linear time series models
through multi-step forecasts. The FCAR model performed much better than
the other two models in terms of bias. However, the FCAR model only
performed better for short term forecasts in terms of mean square
error (MSE). For long term forecasts, the linear model performed the
best in MSE. 

\citet*{cai2000} used a local linear fitting method to estimate $m_{\alpha}\left(\cdot\right)$ in \eqref{eq:fcar1}. They used  the method on simulated data from an EXPAR model and assessed the fit by calculating the square root of  the average squared errors (RASE). The performance of the method was gauged by comparing the RASE to the standard deviation of the time series. Their results showed the local linear method provided an adequate fit of the model with the RASE well below the standard deviation of the series. Real data examples were used to assess the post sample forecasting performance of the local linear method. The two examples were the Canadian lynx data set \citep[p. 377]{tong1990} and the Wolf's annual sunspot numbers data set \citep[p. 420]{tong1990}.   The local linear method  was  compared
with the linear AR model, the TAR model, and the arranged local regression procedure of \citet{chen1993} using a one-step ahead and a iterative two-step ahead forecast. In terms of average absolute predictive errors, the local linear method had much better performance than both the linear AR model and the TAR model in the Canadian Lynx example and performed just as well as the other two models in the sunspot numbers example. 

\cite{huang2004} propose a global smoothing procedure based on polynomial splines for estimating FCAR models. The authors note that the spline method yields a fitted model with a parsimonious explicit expression which is an advantage over the local polynomial method. This feature allows one to produce multi-step ahead forecasts conveniently. Additionally, their spline method is less computationally intensive than the local polynomial method. 

Forecasting for FCAR models was discussed in \cite{harvill2005}. They compared three methods, the first of which is the naive forecasts presented in \cite{fan2003}. Another method was the multistage method of \cite{chen1996}. This method was developed for a general non-linear AR model and was adapted for the FCAR model by \cite{harvill2005}. The last method was the bootstrapping method of \cite{huang2004}. In that paper, bootstrapped residuals are added to the forecasted values after fitting the model with splines. \cite{harvill2005} used bootstrapped residuals after fitting with the local linear method.
A comparison of the three methods showed that the bootstrap method out performs the other two methods for non-linear forecasting and performs well for forecasting a linear process. 

A recent development in estimating nonlinear time series data is the spline-backfitted kernel (SBK)  method of \cite{wang2007}. This method combines the computational speed of splines with the asymptotic properties of kernel smoothing. To estimate a component function in the model, all other component functions are ``pre-estimated" with splines and then the difference is taken of the observed time series and the pre-estimates. This difference is then used as pseudo-responses for which kernel smoothing is used to estimate the function of interest. By constructing the estimates in this way, the method does not suffer from the ``curse of dimensionality." 
The SBK method is adapted for i.i.d.~data in \cite{wang2009}, to generalized additive models in \cite*{liu2011}, and to partially linear additive models in \cite{ma2011}. In \cite{song2010}, a spline-backfitted spline (SBS) procedure is proposed.  \cite{liu2010} develops the SBK method for additive coefficient models which are generalized forms of FCAR models.

Using SBK\ in forecasting algorithms has not been introduced, and in particular, not in forecasting with the FCAR model.  In this paper, we develop new forecasting
algorithms that make use of the SBK\ method, and compare the performance of the new method to that of the naive, the bootstrap, and the multistage methods for the FCAR model.

In Section~\ref{sec:estimation}, the SBK method for estimating the functional coefficients is given.  Section~\ref{sec:forecasting} presents three forecasting methods that employ the SBK estimates. In Section~\ref{sec:simulation}, simulation results are used to compare the forecasting methods and the methods are applied to solar irradiance data in Section~\ref{sec:application}. We conclude with a discussion in Section~\ref{sec:discussion}.

\section{Estimation of Functional Coefficients}
\label{sec:estimation}
To estimate the functional coefficients $m_\alpha(U_t), \alpha = 1, 2, \ldots, p$ in~\eqref{eq:fcar1}, we use the oracle smoothing
of \cite{linton1997} and \cite{wang2007}.  For the remainder of this paper, we take the variable $U_t$ in equation~\eqref{eq:fcar1} to be a lagged value of the series; that is, $U_t = X_{t-d}$, where $d$ is a positive integer.  Consider a fixed integer $\gamma,~1 \le \gamma \le p$. If the coefficient functions $m_{\alpha}\left(X_{t-d}\right)$,
$\alpha=1,\ldots,p$, for $\alpha\ne\gamma$, are known by ``oracle,''
then we can construct $\left\{ X_{t-d},X_{t-\gamma},Y_{\gamma,t}\right\} _{t=1}^{n}$,
where
\[
Y_{\gamma,t}=m_{\gamma}\left(X_{t-d}\right)X_{t-\gamma}+\sigma\left(\textbf{X}_{t}\right)\varepsilon_{t}=X_{t}-\sum_{\alpha=1,\alpha\ne\gamma}^{p}m_{\alpha}\left(X_{t-d}\right)X_{t-\alpha},
\]
from which we can estimate the unknown $m_{\gamma}\left(X_{t-d}\right)$.
This oracle smoother removes the ``curse of dimensionality,'' since
there is only one function being estimated. Clearly, the coefficient
functions, $m_{\alpha}\left(X_{t-d}\right)$, $\alpha=1,\ldots,p$,
$\alpha\ne\gamma$, are not known and must be estimated. For additive
models, \cite{linton1997} used marginal integration kernel estimates to estimate
the functions and \cite{wang2007} used an under-smoothed spline procedure. We
now adapt the procedure of \cite{wang2007} to estimate the FCAR model.

We first ``pre-estimate'' the coefficient functions with a constant spline procedure. Let $U_{t}=X_{t-d}$ be distributed on the compact interval $\left[a,b\right]$ and denote the knots as $a=\kappa_{0}<\kappa_{1}<\cdots<\kappa_{N}<\kappa_{N+1}=b$
where the number of interior knots are $N\sim n^{1/4}\ln n$. The
B-spline basis functions are determined on the $N+1$ equally spaced
intervals with length $(b-a)\left(N+1\right)^{-1}$. The basis function
are defined as
\[
B_{J}\left(u\right)=\begin{cases}
1, & \qquad\kappa_{J}\le x<\kappa_{J+1},\\
0, & \qquad\text{otherwise,}
\end{cases}\qquad J=0,\ldots,N.
\]
The pre-estimates are defined as
\[
\hat{m}_{\alpha}\left(u\right)=\sum_{J=0}^{N}\hat{\lambda}_{\left(N+1\right)(\alpha-1)+J}B_{J}\left(u\right),\qquad\alpha=1,\ldots,p,
\]
where the coefficients $(\hat{\lambda}_{0},\ldots,\hat{\lambda}_{p\left(N+1\right)-1})$
are solutions to the least squares problem
\begin{align}
&\left\{ \hat{\lambda}_{0},\ldots, \hat{\lambda}_{p\left(N+1\right)-1}\right\} = \nonumber \\  &\underset{\mathbb{R}^{p\left(N+1\right)-1}}{\arg\min}\sum_{t=1}^{n}\left\{ X_{t}-\sum_{\alpha=1}^{p}\left(\sum_{J=0}^{N}\lambda_{\left(N+1\right)(\alpha-1)+J}B_{J}\left(U_{t}\right)\right)X_{t-\alpha}\right\} ^{2}.
\label{eq:leastsquares}
\end{align}
Define $\mathbf{X}_{t}=\left(X_{1},\ldots,X_{n}\right)^{\prime}$,
$\mathbf{X}_{\alpha}=\left(X_{1-\alpha},\ldots,X_{n-\alpha}\right)^{\prime},$
$\mathbf{U}_{t}=\left(U_{1},\ldots,U_{n}\right)^{\prime}$, $\hat{\boldsymbol{\lambda}}=(\hat{\lambda}_{0},\ldots,\hat{\lambda}_{p\left(N+1\right)-1})^{\prime}$,
\[
\mathbf{B}=\left[\begin{array}{cccc}
B_{0}\left(U_{1}\right) & B_{1}\left(U_{1}\right) & \cdots & B_{N}\left(U_{1}\right)\\
B_{0}\left(U_{2}\right) & B_{1}\left(U_{2}\right) & \cdots & B_{N}\left(U_{2}\right)\\
\vdots & \vdots & \ddots & \vdots\\
B_{0}\left(U_{n}\right) & B_{1}\left(U_{n}\right) & \cdots & B_{N}\left(U_{n}\right)
\end{array}\right],
\]
and $\mathbf{Z}=(\mathbf{B}\circ\tilde{\mathbf{X}}_{1},\mathbf{B}\circ\tilde{\mathbf{X}}_{2},\cdots,\mathbf{B}\circ\tilde{\mathbf{X}}_{p})$
where $\circ$ denotes the Hadamard product and $\tilde{\mathbf{X}}_{\alpha}$
is a $n\times\left(N+1\right)$ matrix with $\mathbf{X}_{\alpha}$
for each column. In matrix notation, the least squares estimates are
\[
\hat{\boldsymbol{\lambda}}=\left(\mathbf{Z}^{\prime}\mathbf{Z}\right)^{-1}\mathbf{Z}^{\prime}\mathbf{X}_{t},
\]
and the pre-estimates are 
\[
\hat{m}_{\alpha}\left(\mathbf{U}_{t}\right)=\mathbf{B}\left[\hat{\lambda}_{\left(N+1\right)\left(\alpha-1\right)},\ldots,\hat{\lambda}_{\alpha\left(N+1\right)-1}\right]^{\prime},\qquad\alpha=1,\ldots,p.
\]
We now define the ``pseudo-responses'' as
\[
\hat{Y}_{\gamma,t}=X_{t}-\sum_{\alpha=1,\alpha\ne\gamma}^{p}\hat{m}_{\alpha}\left(U_{t}\right)X_{t-\alpha},\qquad t=1,\ldots n.
\]
Define the vector of pseudo-responses as $\hat{\mathbf{Y}}_{\gamma}=(\hat{Y}_{\gamma,1},\ldots,\hat{Y}_{\gamma,n})^{\prime}$.
The spline-backfitted local linear (SBLL) estimate for the coefficient function
$m_{\gamma}\left(u\right)$ is 
\begin{equation}
\tilde{m}_{SBLL,\gamma}\left(u\right)=\left(1,0\right)\left(\mathbf{V}^{\prime}\mathbf{W}\mathbf{V}\right)^{-1}\mathbf{V}^{\prime}\mathbf{W}\hat{\mathbf{Y}}_{\gamma},
\label{eq:sbll}
\end{equation}
where 
\[
\mathbf{V}=\left[\begin{array}{cc}
X_{p+1-\gamma} & X_{p+1-\gamma}\left(U_{p+1}-u\right)\\
\vdots & \vdots\\
X_{n-\gamma} & X_{n-\gamma}\left(U_{n}-u\right)
\end{array}\right],
\]
$\mathbf{W}=\text{diag}\left\{ K_{h}\left(U_{p+1}-u\right),\ldots,K_{h}\left(U_{n}-u\right)\right\} $,
\[
K_{h}\left(u\right)=h^{-1}\frac{15}{16}\left\{1-\left(\frac{u}{h}\right)^{2}\right\}^{2}I_{\left\{ \left|u/h\right|\le1\right\}},
\]
$I_{\{A\} }$ is an indicator variable equal to one if
condition $A$ is true, and is zero otherwise; $h$ is a bandwidth selected by the rule of thumb criterion of \cite{fan1996}. 
Likewise, we define the oracle local linear smoother as
\[
\tilde{m}_{O,\gamma}\left(u\right)=\left(\mathbf{V}^{\prime}\mathbf{W}\mathbf{V}\right)^{-1}\mathbf{V}^{\prime}\mathbf{W}\mathbf{Y}_{\gamma},
\]
\cite{wang2007} first proposed the method using a Nadaraya-Watson smoother in the last step and then proposed using a local linear smoother.  We use only the local linear smoother in this paper.

\subsection{Asymptotic properties}
\label{sec:asymptotics}
The following assumptions are necessary for the data generating process to be geometric ergodic and for the asymptotic properties of the SBLL method.
%
%
For the interval $\left[a,b\right]$ and functions $m$, let $C^{\left(2\right)}\left[a,b\right]=\left\{ m|m^{\prime\prime}\in C\left[a,b\right]\right\} $
denote the space of second-order continuous smooth functions and let 
$$
\text{Lip}\left(\left[a,b\right],C\right)=\left\{ m\mid\left|m\left(u\right)-m\left(v\right)\right|\le C\left|u-v\right|,\forall u,v\in\left[a,b\right]\right\} 
$$
denote the class of Lipschitz continuous functions for any fixed constant $C>0$. 
\begin{enumerate}
\item[(A1)]\label{a1} The coefficient function $m_{\gamma}\left(u\right)\in C^{\left(2\right)}\left[a,b\right]$
and $m_{\alpha}\left(u\right)\in\text{Lip}\left(\left[a,b\right],C_{\infty}\right)$,
$\alpha=1,\ldots,p$, $\alpha\ne\gamma$, for some constant $0<C_{\infty}<\infty$. 
\item[(A2)] For the process $\left\{ \boldsymbol{\theta}_{t}=\left(U_{t},X_{t},X_{t-1},\ldots X_{t-p},\varepsilon_{t}\right)\right\} _{t=1}^{n}$,
there exist positive constants $K_{0}$ and $\lambda_{0}$ such that
$\alpha\left(k\right)\le K_{0}e^{-\lambda_{0}k}$ holds for all $k$,
with the $\alpha$-mixing coefficients for $\left\{ \theta_{t}\right\} _{t=1}^{n}$
defined as
\[
\alpha\left(k\right)=\sup_{B\in\sigma\left\{ \boldsymbol{\theta}_{s},s\le t\right\} ,C\in\sigma\left\{ \boldsymbol{\theta}_{s}\ge t+k\right\} }\left|P\left(B\cap C\right)-P\left(B\right)P\left(C\right)\right|,\qquad k\ge1.
\]
\item[(A3)] The conditional variance function $\sigma^{2}\left(\mathbf{X}_{t}\right)$
is measurable and bounded. The noise $\varepsilon_{t}$ satisfies
$E\left(\varepsilon_{t}|\mathbf{X}_{t}\right)=0$, $E\left(\varepsilon_{t}^{2}|\mathbf{X}_{t}\right)=1$
and $E\left(\left|\varepsilon_{t}\right|^{2+\delta}|\mathbf{X}_{t}\right)<M_{\delta}$
for some $\delta>1/2$ and a finite positive $M_{\delta}$.
\item[(A4)] The delay variable $U_{t}$ has a continuous probability density function
$f\left(u\right)$ that satisfies 
\[
0<c_{f}\le\inf_{u\in\left[a,b\right]}f\left(u\right)\le\sup_{u\in\left[a,b\right]}f\left(u\right)\le C_{f}<\infty,
\]
 for some constants $c_{f}$ and $C_{f}$, and has continuous derivatives
on $\left[a,b\right]$.
\item[(A5)] The kernel function $K\in\text{Lip}\left(\left[-1,1\right],C_{k}\right)$
for some constant $C_{k}>0$, and is bounded, nonnegative, symmetric
and supported on $\left[-1,1\right]$. The bandwidth is $c_{h}n^{-1/5}\le h\le C_{h}n^{-1/5}$
for some positive constants $c_{h}$ and $C_{h}$. 
\item[(A6)] The number of interior knots is $c_{N}n^{1/4}\log n\le N\le C_{N}n^{1/4}\log n$
for some positive constants $c_{N}$ and $C_{N}$.
\end{enumerate}

Assumptions (A1)-(A5) are common in the nonparametric literature;
see, for example \cite{fan1996}. We apply the SBLL method to a TAR model in section \ref{sec:simulation}, thus relaxing the smoothness assumption in assumption (A1). Theorem 1.1 in \cite{chen1993} gives sufficient conditions for FCAR
models to be geometrically ergodic, which implies $\alpha$-mixing,
thus satisfying Assumption (A2). Assumption (A6) is used in the pre-estimate
stage in which we under smooth with the splines to reduce the bias.
The increase in variance resulting from the pre-estimate stage is
reduced in the local linear regression stage where bandwidth $h$
is of the order found in assumption (A5). When implementing the method,
we impose an additional constraint on $N$ such that 
\[
N=\min\left\{ n^{1/4}\log n,\left(\frac{n}{2p}\right)-1\right\} .
\]
This additional constraint ensures that the number of terms in the
least squares problem \eqref{eq:leastsquares} is no greater than $n/2$.

To simplify the notation, we denote
\[
\mu_{j}=\int_{-\infty}^{\infty}x^{j}K\left(x\right)dx,\qquad v_{j}=\int_{-\infty}^{\infty}x^{j}K^{2}\left(x\right)dx,
\]
\[
\boldsymbol{\Omega}_{\gamma}\left(u\right)=E\left(\textbf{X}_{\gamma}\textbf{X}_{\gamma}^{\prime}\mid U_{t}=u\right),
\]
and
\[
\boldsymbol{\Omega}_{\gamma}^{*}\left(u\right)=E\left(\textbf{X}_{\gamma}\textbf{X}_{\gamma}^{\prime}\sigma^{2}\left(U_{t},\textbf{X}_{t}\right)\mid U_{t}=u\right).
\]
Under assumptions (A1)-(A5), it is straight forward from the results
of \cite{cai2000} to verify that, as $n\rightarrow\infty$, 
\[
\sqrt{nh}\left\{ \tilde{m}_{O,\gamma}\left(u\right)-m_{\gamma}\left(u\right)-b_{\gamma}\left(u\right)h^{2}\right\} \rightarrow N\left(0,v_{\gamma}^{2}\left(u\right)\right),
\]
where 
\begin{equation}
b_{\gamma}\left(u\right)=\frac{h^{2}}{2}\mu_{2}m_{\gamma}^{\prime\prime}\left(u\right),
\label{eq:bias}
\end{equation}
\begin{equation}
v_{\gamma}^{2}\left(u\right)=\frac{\nu_{0}}{f\left(u\right)}\textbf{e}_{\gamma,p}^{\prime}\boldsymbol{\Omega}^{-1}\left(u\right)\boldsymbol{\Omega}^{*}\left(u\right)\boldsymbol{\Omega}^{-1}\left(u\right)\textbf{e}_{\gamma,p},
\label{eq:variance}
\end{equation}
and $\textbf{e}_{j,p}$ is a $p\times1$ vector with 1 in the $j$th
position. Furthermore, from the results of \cite{liu2010} it can be shown
that as $n\rightarrow\infty$, the oracle smoother satisfies 
\[
\sup_{u\in\left[a+h,b-h\right]}\left|\tilde{m}_{O,\gamma}\left(u\right)-m_{\gamma}\left(u\right)\right|=O_{p}\left(\frac{\log n}{\sqrt{nh}}\right).
\]
The following theorem gives the asymptotic uniform magnitude of the
diffrence between $\tilde{m}_{SBLL,\gamma}\left(u\right)$ and $\tilde{m}_{O,\gamma}\left(u\right)$.
\begin{theorem}
\label{th:1}
Under assumptions (A1)-(A6), as $n\rightarrow\infty$, the SBLL estimator
$\tilde{m}_{SBLL,\gamma}\left(u\right)$ given in \eqref{eq:sbll} satisfies
\[
\sup_{u\in\left[a,b\right]}\left|\tilde{m}_{SBLL,\gamma}\left(u\right)-\tilde{m}_{O,\gamma}\left(u\right)\right|=o_{p}\left(n^{-2/5}\right).
\]
\end{theorem}
Theorem \ref{th:1} states that the distance $\tilde{m}_{SBLL,\gamma}\left(u\right)-\tilde{m}_{O,\gamma}\left(u\right)$
is of the order $o_{p}\left(n^{-2/5}\right)$, which is dominated
by the asymptotic size of $\tilde{m}_{O,\gamma}\left(u\right)-m_{\gamma}\left(u\right)$.
This implies that $\tilde{m}_{SBLL,\gamma}\left(u\right)$ will have
the same asymptotic distribution as $\tilde{m}_{O,\gamma}\left(u\right)$
which results in the following theorem.
\begin{theorem}
\label{th:2}
Under assumptions (A1)-(A6), as $n\rightarrow\infty$, with $b_{\gamma}\left(u\right)$
and $v_{\gamma}^{2}\left(u\right)$ defined in \eqref{eq:bias} and \eqref{eq:variance}, 

\[
\sqrt{nh}\left\{ \tilde{m}_{SBLL,\gamma}\left(u\right)-m_{\gamma}\left(u\right)-b_{\gamma}\left(u\right)h^{2}\right\} \rightarrow N\left(0,v_{\gamma}^{2}\left(u\right)\right).
\]
\end{theorem}
\noindent
The proofs of Theorems \ref{th:1} and \ref{th:2} follow from the proofs of \cite{liu2010}. 

To compare the performance of $\tilde{m}_{SBLL,\gamma}\left(u\right)$
to $\tilde{m}_{O,\gamma}\left(u\right)$, we use the oracle efficiencies
of \cite{wang2007} which are defined as
\begin{equation}
\label{eq:eff}
\text{eff}_{\gamma}=\left[\frac{\sum_{t=1}^{n}\left\{ \tilde{m}_{O,\gamma}\left(u_t\right)-m_{\gamma}\left(u_t\right)\right\} ^{2}}{\sum_{t=1}^{n}\left\{ \tilde{m}_{SBLL,\gamma}\left(u_t\right)-m_{\gamma}\left(u_t\right)\right\} ^{2}}\right]^{1/2}.
\end{equation}
By Theorems \ref{th:1} and \ref{th:2}, $\text{eff}_{\gamma}$ should approach
1 as $n\rightarrow\infty$ for all $\gamma=1,\ldots,p$. We demonstrate this result via simulation in Section \ref{sec:simulation}.

\section{Forecasting Methods}
\label{sec:forecasting}
We now present the three forecasting methods discussed in \cite{harvill2005} and adapt them to be used with the SBLL method. 
Assuming $m_{\alpha}\left(U\right)$ is known and $U$ is exogenous, we want to find an estimator of the conditional expectation
\begin{align}
E\left[X_{n+M}\vert X_{n},\ldots,X_{n-p}\right] & =  E\left[\sum_{\alpha=1}^{p}m_{\alpha}\left(U_{n+M}\right)X_{n+M-\alpha}\vert X_{n},\ldots,X_{n-p}\right]\label{eq:expect}\\
 & =  \sum_{\alpha=1}^{p}m_{\alpha}\left(U_{n+M}\right)E\left[X_{n+M-\alpha}\vert X_{n},\ldots,X_{n-p}\right]\nonumber\\
 & =  \sum_{\alpha=1}^{p}m_{\alpha}\left(U_{n+M}\right)\hat{X}_{n+M-\alpha}.\nonumber
 \end{align}
The expectation in \eqref{eq:expect} is no longer a simple linear operation when
$U_{t}=X_{t-d}$ for some positive constant $d$. The three forecasting methods described below deal with this expectation in a different way.

\subsubsection*{Naive predictor}

The naive approach simply ignores  the fact the expectation in \eqref{eq:expect} is not a linear function of $X_{t+M-\alpha}$ and  substitutes  $\hat{X}_{t+M-\alpha}$ into the forecast equation. We estimate the coefficient function only using the within-sample series values. The naive predictor is defined as 
\begin{align*}
\hat{X}_{n+M} & =\sum_{\alpha=1}^{p}\hat{m_\alpha}\left(\hat{X}_{n+M-d}\right)\hat{X}_{n+M-\alpha},\end{align*}
where $\hat{X}_{t}=X_{t}$, $t\le n$. For the SBLL estimator,  $\hat{m}_\alpha\left(\cdot\right)$ is the value obtained by the general form of \eqref{eq:sbll}. Thus, the spline pre-estimate   is not computed for each value of $\hat{X}_{n+M-d}$, but the local linear estimation is computed. 

\subsubsection*{Bootstrap predictor}

The bootstrap predictor is like the naive predictor in that it estimates the functional coefficients using only the within-sample values. However, we bootstrap the within-sample residuals from the estimated model and find the predicted value as 
\begin{align*}
\hat{X}_{n+M} & =\sum_{\alpha=1}^{p}\hat{m}_{\alpha}\left(\hat{X}_{n+M-d}\right)\hat{X}_{n+M-\alpha}+\epsilon^{b},
\end{align*}
where $\epsilon^{b}$ is the bootstrapped residual. We obtain bootstrapped
forecasts for $b=1,\ldots,B$, and use the average of these values
as the $M$-step ahead forecast. For the SBLL method, we estimate $\hat{m}_{\alpha}\left(\cdot\right)$
as with the naive predictor. An advantage of using the bootstrap
values is that the set of all  values allows us to estimate
the predictive density of $X_{t+M}$. A disadvantage is that the estimated
functional coefficients may become unreliable when $\hat{X}_{t+M-d}$
is outside or near the boundary of the range of the original $X_{t-d}$.
This disadvantage was first noted by \cite{huang2004} and reiterated by \cite{harvill2005}. 

\subsubsection*{Multistage predictor}

Another way to handle  the expectation in \eqref{eq:expect} is to incorporate the information from $X_t$ encoded in the predicted response at time $n+j,~j=1,\ldots,M-1$. This is accomplished
by updating the functional coefficients at each step and obtaining the forecast by
\begin{align*}
\hat{X}_{n+M} & =\sum_{\alpha=1}^{p}\hat{m}_{\alpha}^{M}\left(\hat{X}_{n+M-d}\right)\hat{X}_{n+M-\alpha},
\end{align*}
where $\hat{X}_{t}=X_{t}$, $t\le n$. The functional coefficient
$\hat{m}_{\alpha}^{M}\left(\cdot\right)$ is estimated by the SBLL method
at each step. That is, we include the predicted values $\hat{X}_{t}$,
$t=n+1,\ldots,M-1,$ with the original values $X_{t}$, $t=1,\ldots,n$,
and then re-estimate the functional coefficient with the SBLL method
using the new set of data. Clearly, this method is more computationally intensive than the naive predictor and possibly the bootstrap predictor, dependent on the number of bootstraps taken.

\section{Simulation}
\label{sec:simulation}
In the following, we investigate the performance of the three forecasting methods via simulation.
We use three parametric models, described in more detail in each of the three examples.  An FCAR model
is fit to data from the three models using the method described in Section~\ref{sec:estimation}, and forecasts obtained using the methods explained in Section~\ref{sec:forecasting}.  The models in the first two examples satisfy all assumptions provided in Section~\ref{sec:asymptotics}.  However, the model in the third example is a self-exciting threshold autoregressive model, and so the continuity assumption (A1) is not satisfied.  For all three models, we provide empirical efficiencies as defined in equation~\eqref{eq:eff}.  We also provide the root mean square prediction error (RMPE) for the three forecasting methods.  For each model, we ran 500 Monte Carlo iterations for series lengths $n = 75, 150, 250$ and 500, and orders $p = 4, 10$.


\begin{example}
\label{ex:sine}
We first consider the model 
\[
X_{t}=\sum_{\alpha=1}^{p}a_{\alpha}\sin\left(\omega\pi X_{t-d}\right)X_{t-\alpha}+\sigma\left(\textbf{X}_t\right)\varepsilon_{t},\qquad\varepsilon_{t}\sim N\left(0,1\right),
\]
where
\begin{equation}
\label{eq:sigma}
\sigma\left(\textbf{X}_{t}\right)=0.1\left(\frac{\sqrt{p}}{2}\right)\frac{5-\exp\left(\sum_{\alpha=1}^{p}\left|X_{t-\alpha}\right|/p\right)}{5-\exp\left(\sum_{\alpha=1}^{p}\left|X_{t-\alpha}\right|/p\right)}.
\end{equation}
In both cases, we set the delay to $d = 2$. The term $\sigma\left(\textbf{X}_{t}\right)$ ensures the model is
heteroscedastic with the variance roughly proportional to $p$. For
$p=4$, we used the parameters $$\textbf{a}=\left(0.5,-0.5,0.5,-0.5\right)^{\prime}$$
and $\omega=4.5$. For $p=10$, we used $$\textbf{a}=\left(0.5,-0.5,0.5,-0.5,0.5,-0.5,0.5,-0.5,0.5,-0.5\right)^{\prime}$$
and $\omega=1.5$. These choices for $\textbf{a}$ make the
bounds of the coefficient functions $\pm\left|a_{\alpha}\right|$ so
that the roots of the characteristic polynomial 
\[
\lambda^{p}-a_{1}\lambda^{p-1}-\cdots-a_{p}=0
\]
 are inside the unit circle, thus ensuring the process is geometrically
ergodic \citep[see][]{chen1993}. 

Figures \ref{fig:estsinea} -- \ref{fig:estsined} show the estimated coefficient function using the SBLL method and the oracle estimate for $m_{\alpha}\left(X_{t-2}\right)$ with $p=4,10$ and $n=75,500$. The densities of the empirical efficiencies are shown in Figures \ref{fig:estsinee} and \ref{fig:estsinef}. 
%
%
 We can see that the mode of the densities tend to be closer to one as $n$ increases confirming  the convergence results of Theorem \ref{th:1}.

\begin{center}
{\large{Figure~\ref{fig:estsine} about here.}}
\end{center}

For comparing the three forecasting methods, we calculate the root
mean prediction error (RMPE),
\begin{equation}
\label{eq:rmpe}
\text{RMPE}_{M}=\left\{ \frac{1}{500}\sum_{i=1}^{500}\left(\hat{X}_{n+M,i}-X_{n+M,i}\right)^{2}\right\} ^{1/2}\qquad M=1,\ldots10,
\end{equation}
where $\hat{X}_{n+M}$ is the forecast at time $n+M$ for
iteration $i$, and $X_{n+M}$ is the value at time $n+M$ of the
$i$th simulated process. The RMPE is calculated for each of the three
forecasting methods. Figure \ref{fig:rmpesine} shows the $\text{RMPE}_{M}$, $M=1,\ldots10$,
for Example \ref{ex:sine} for three series lengths. For $p=4$, $n=75$ (Figure \ref{fig:rmpesinea}),
we can see that the multistage method has lower RMPE for $M=1,\ldots7$,
but has higher RMPE than the naive method for $M=8,9,10$. For $p=10,$
$n=75$ (Figure \ref{fig:rmpesineb}), the naive and multistage methods are much closer
at the lower values of $M$ with the naive method performing better
for the larger values of $M$. For $p=4$, the larger series lengths
show the naive method having RMPE just as low, if not lower, than
the other two methods (Figures \ref{fig:rmpesinec} and \ref{fig:rmpesinee}). For $p=10$, the larger
series lengths show the multistage method having the lowest RMPE for
some smaller values of $M$ with the naive method being lower for
the larger values of $M$ (Figures \ref{fig:rmpesined} and \ref{fig:rmpesinef}). The simulation results
show that the naive method performs just as well, if not better, than
both the bootstrap and multistage methods. An additional advantage
in using the naive method over the other two is that it is computationally
faster. 

\begin{center}
{\large{Figure~\ref{fig:rmpesine} about here.}}
\end{center}

\end{example}

\begin{figure}[ht]
\begin{center}
\subfigure[]{\label{fig:estsetara}{%
  \includegraphics[height=0.25\textheight]{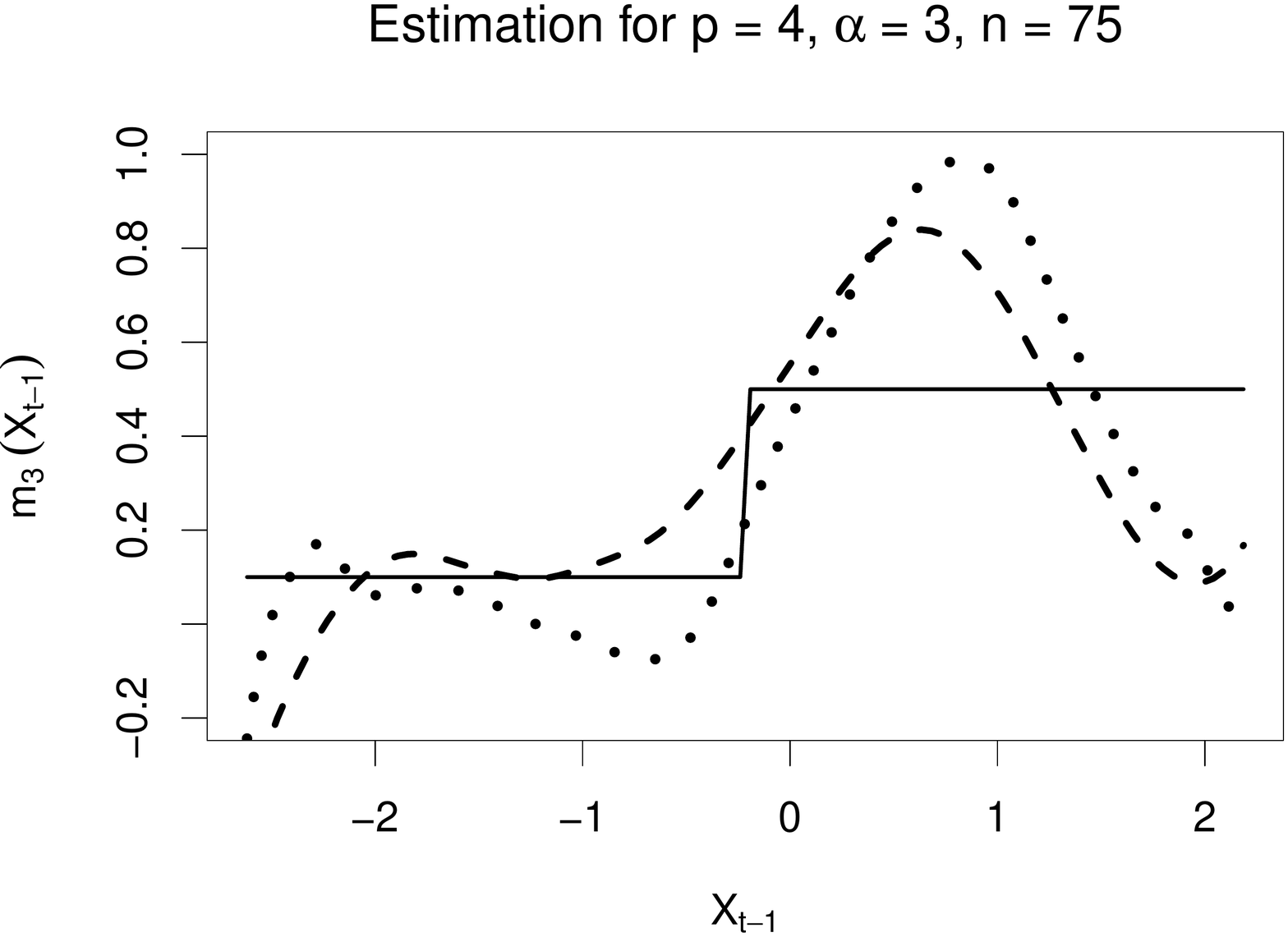}}}
\subfigure[]{\label{fig:estsetarb}{%
  \includegraphics[height=0.25\textheight]{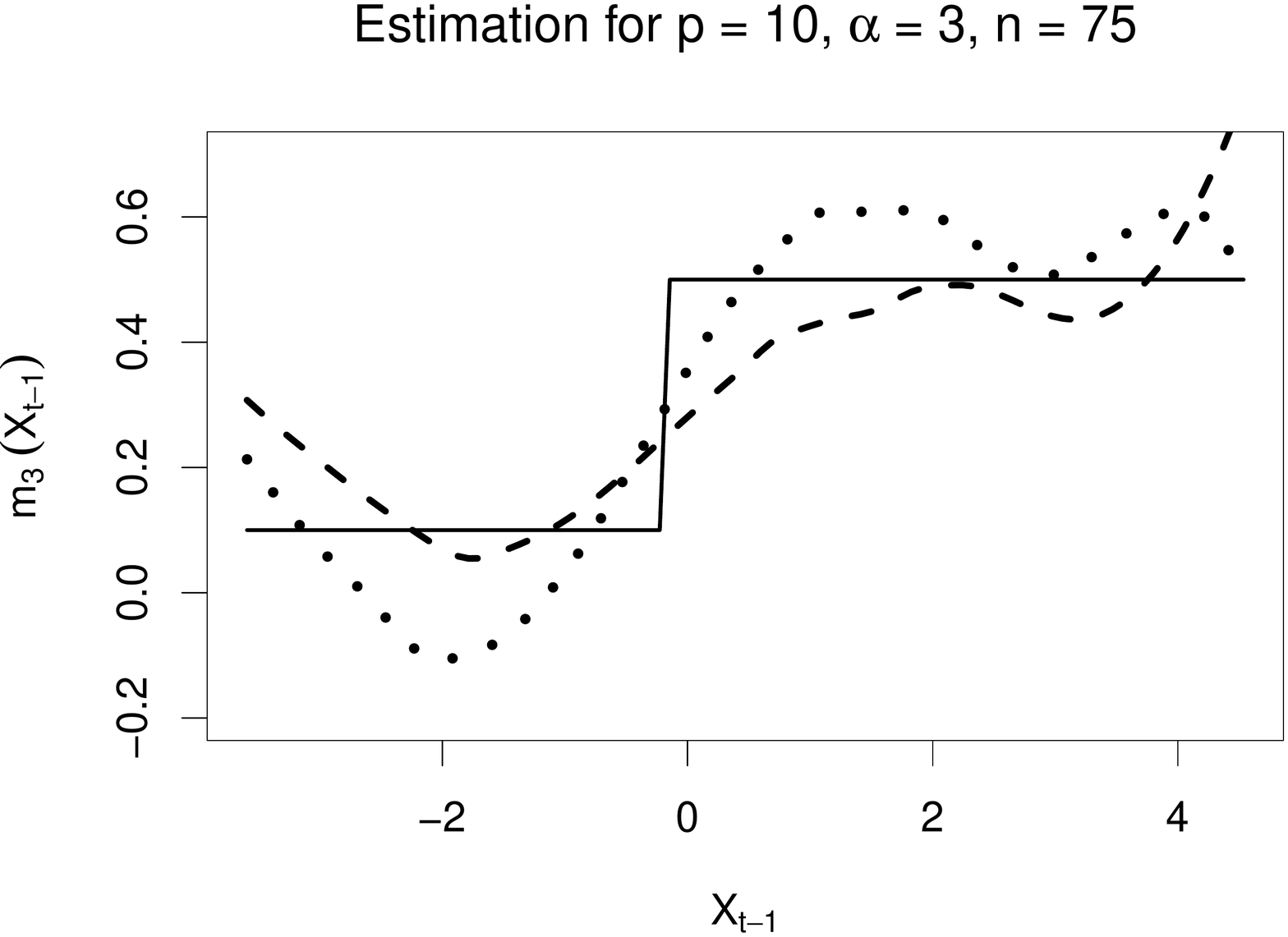}}}
\subfigure[]{\label{fig:estsetarc}{%
  \includegraphics[height=0.25\textheight]{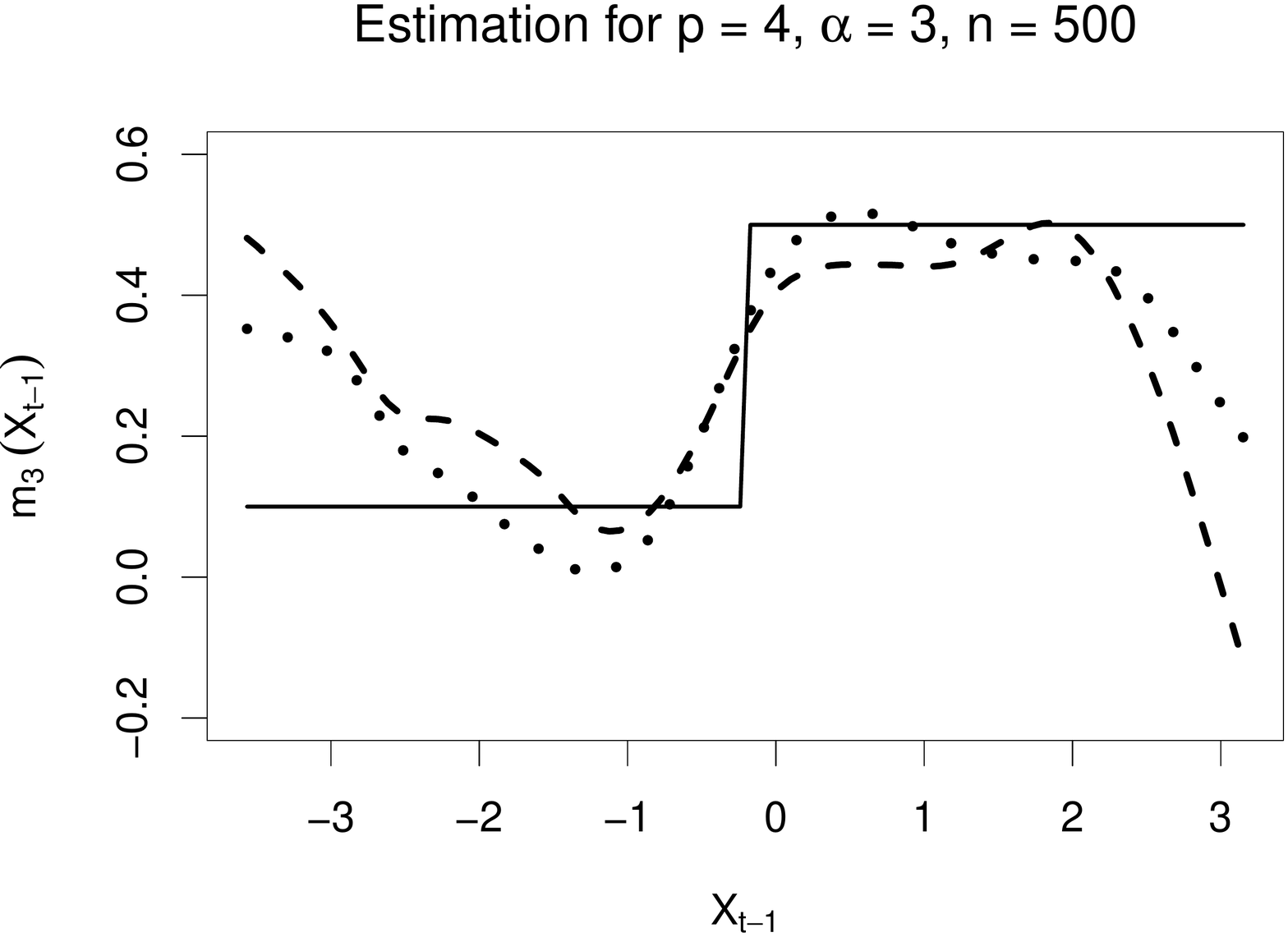}}}
\subfigure[]{\label{fig:estsetard}{%
  \includegraphics[height=0.25\textheight]{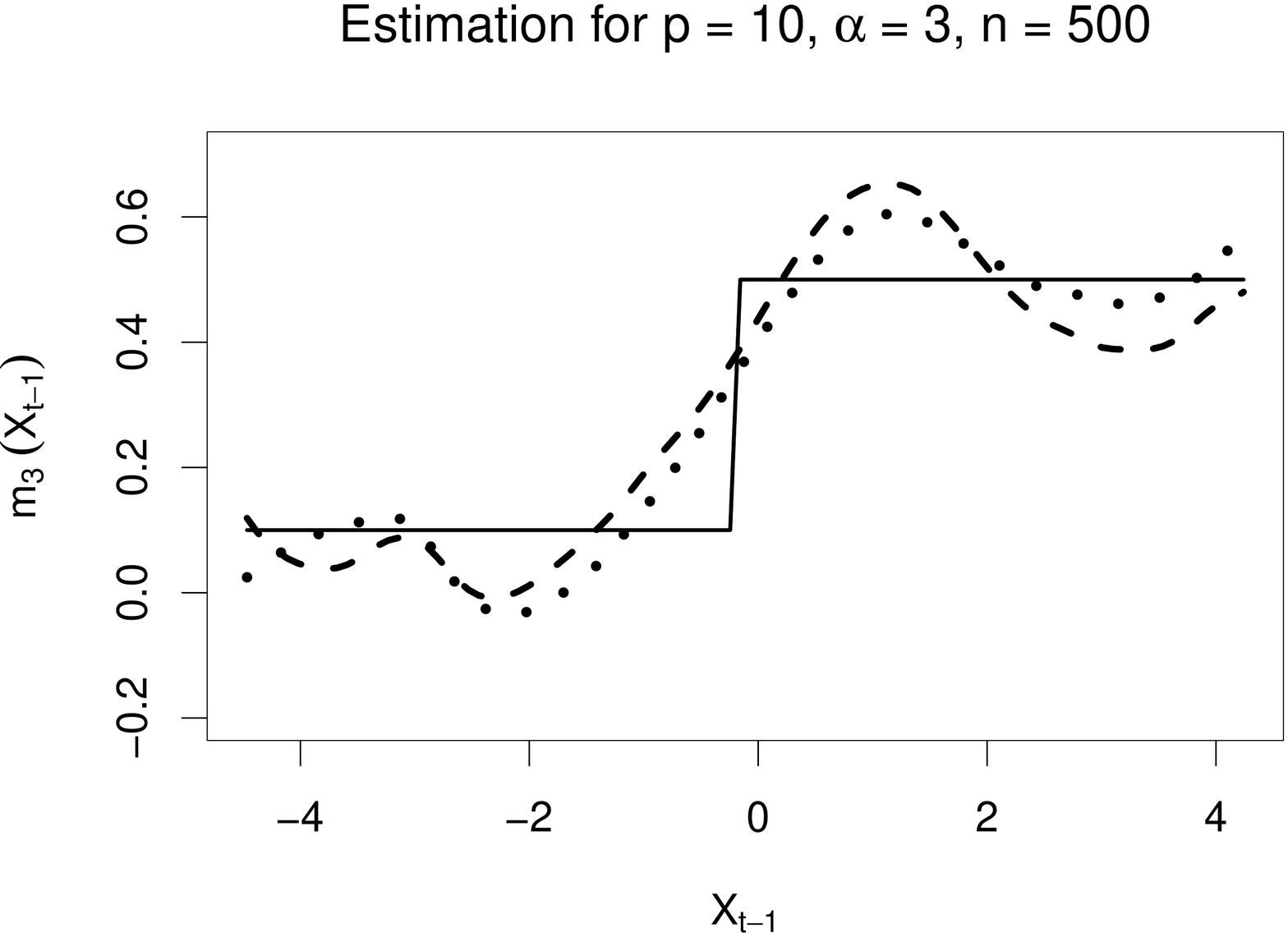}}}
\subfigure[]{\label{fig:estsetare}{%
  \includegraphics[height=0.25\textheight]{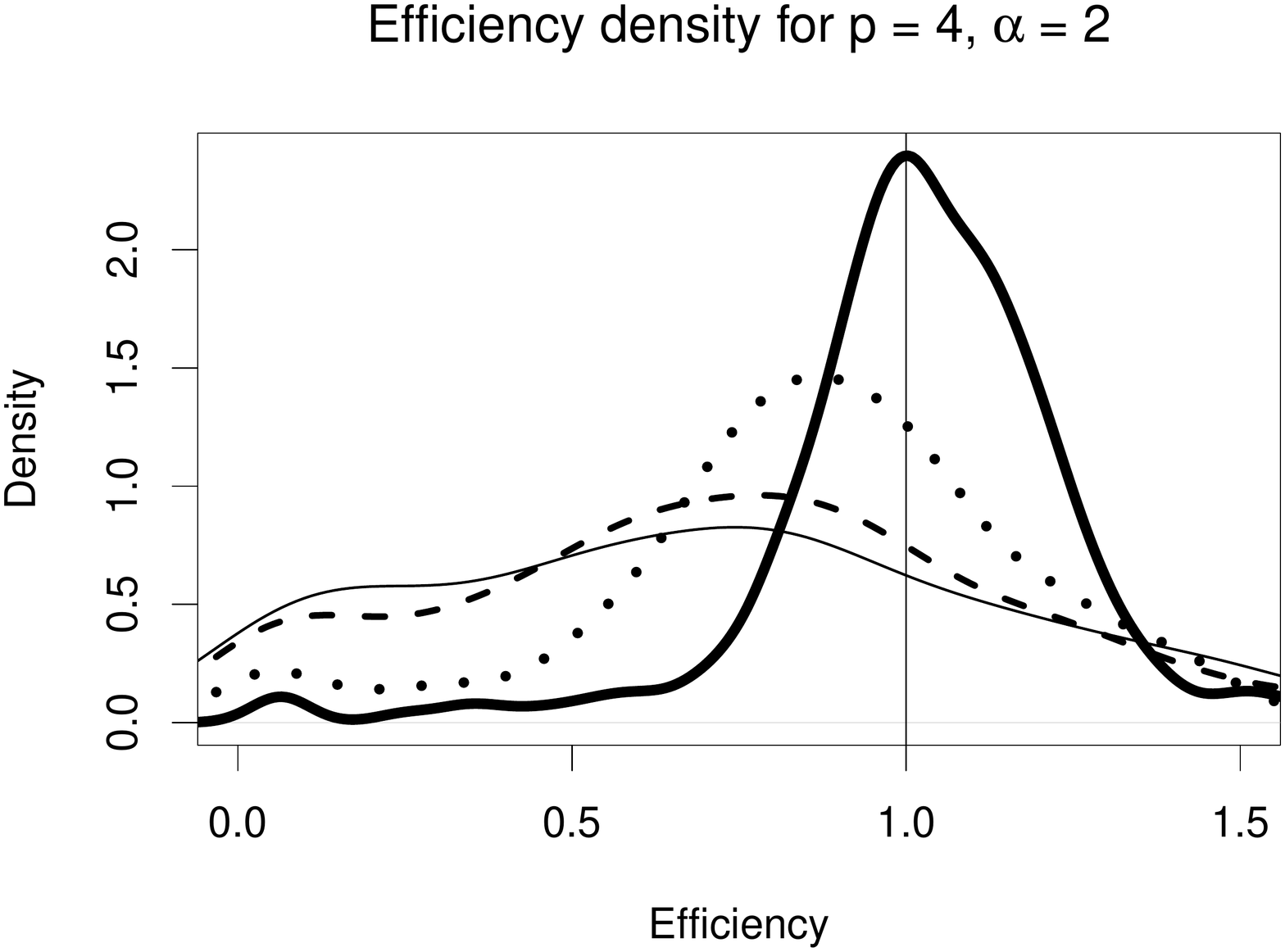}}}
\subfigure[]{\label{fig:estsetarf}{%
  \includegraphics[height=0.25\textheight]{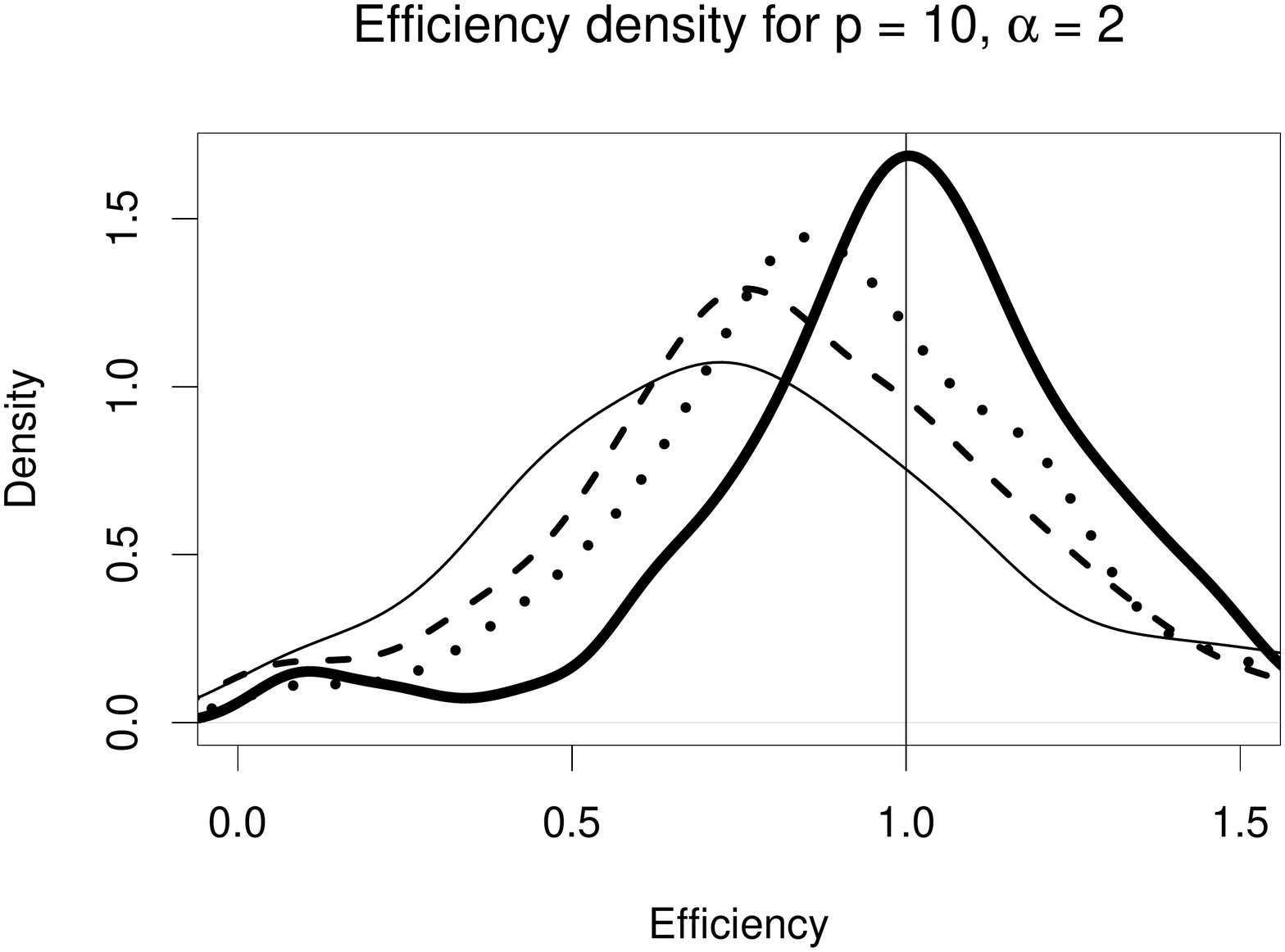}}}
\caption{Plots (a) -- (d) are graphs of the true coefficient function (solid line), the SBLL estimate (dashed line), and the oracle estimate (dotted line) for Example \ref{ex:setar} : (a) $p=4$, $n=75$, (b) $p=10$, $n=75$, (c) $p=4$, $n=500$, (d) $p=10$, $n=500$. Plots (e) and (f) contain the empirical efficiency densities for Example \ref{ex:sine} with series lengths $n=75$ (thin solid line), $150$ (dashed line), $250$ (dotted line), and $500$ (thick solid line) for $p=4$ (e) and $p=10$ (f).}
\label{fig:estsetar}
\end{center}
\end{figure}

\begin{figure}[ht]
\begin{center}
\subfigure[]{\label{fig:rmpesetara}{%
  \includegraphics[height=0.25\textheight]{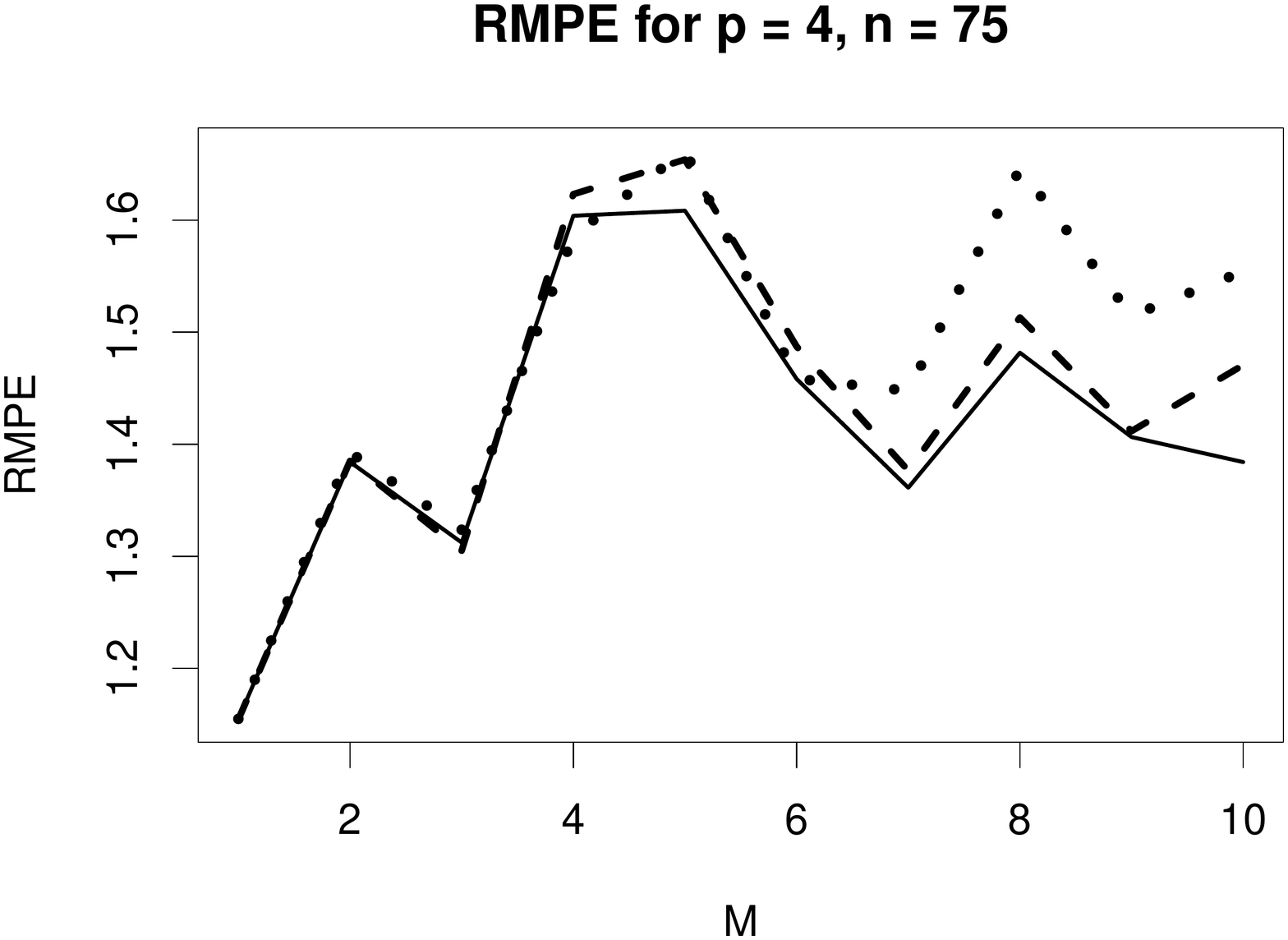}}}
\subfigure[]{\label{fig:rmpesetarb}{%
  \includegraphics[height=0.25\textheight]{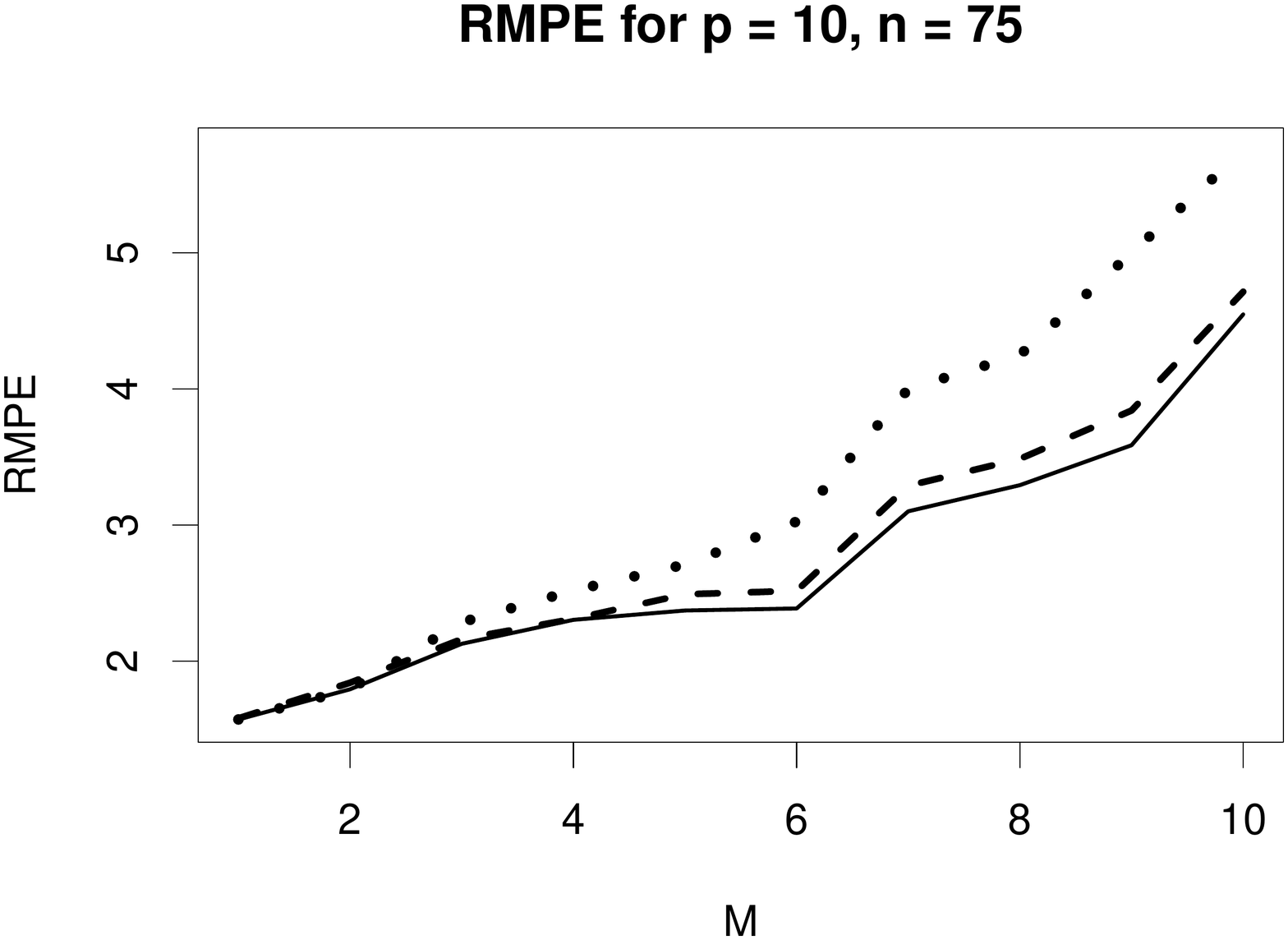}}}
\subfigure[]{\label{fig:rmpesetarc}{%
  \includegraphics[height=0.25\textheight]{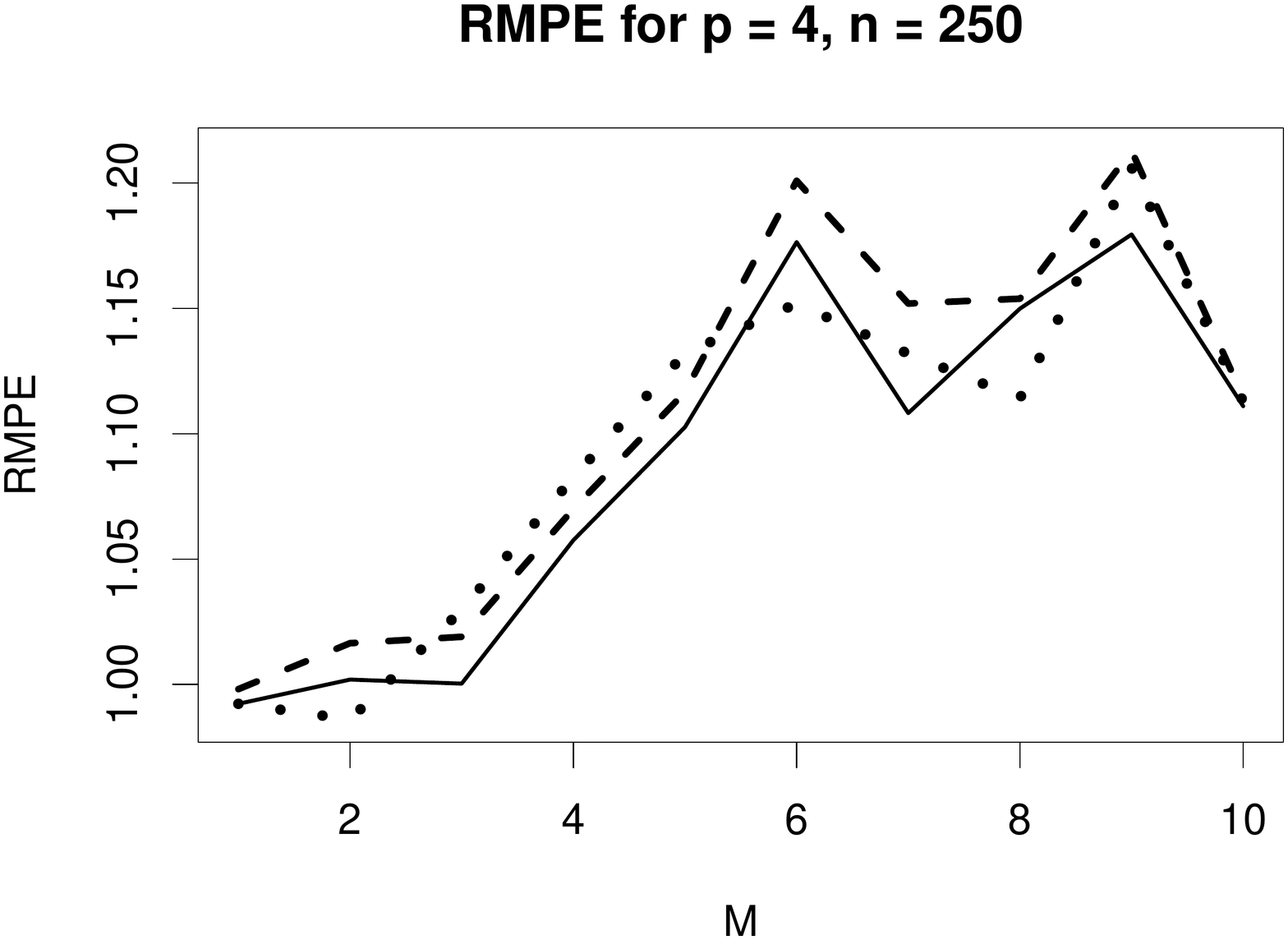}}}
\subfigure[]{\label{fig:rmpesetard}{%
  \includegraphics[height=0.25\textheight]{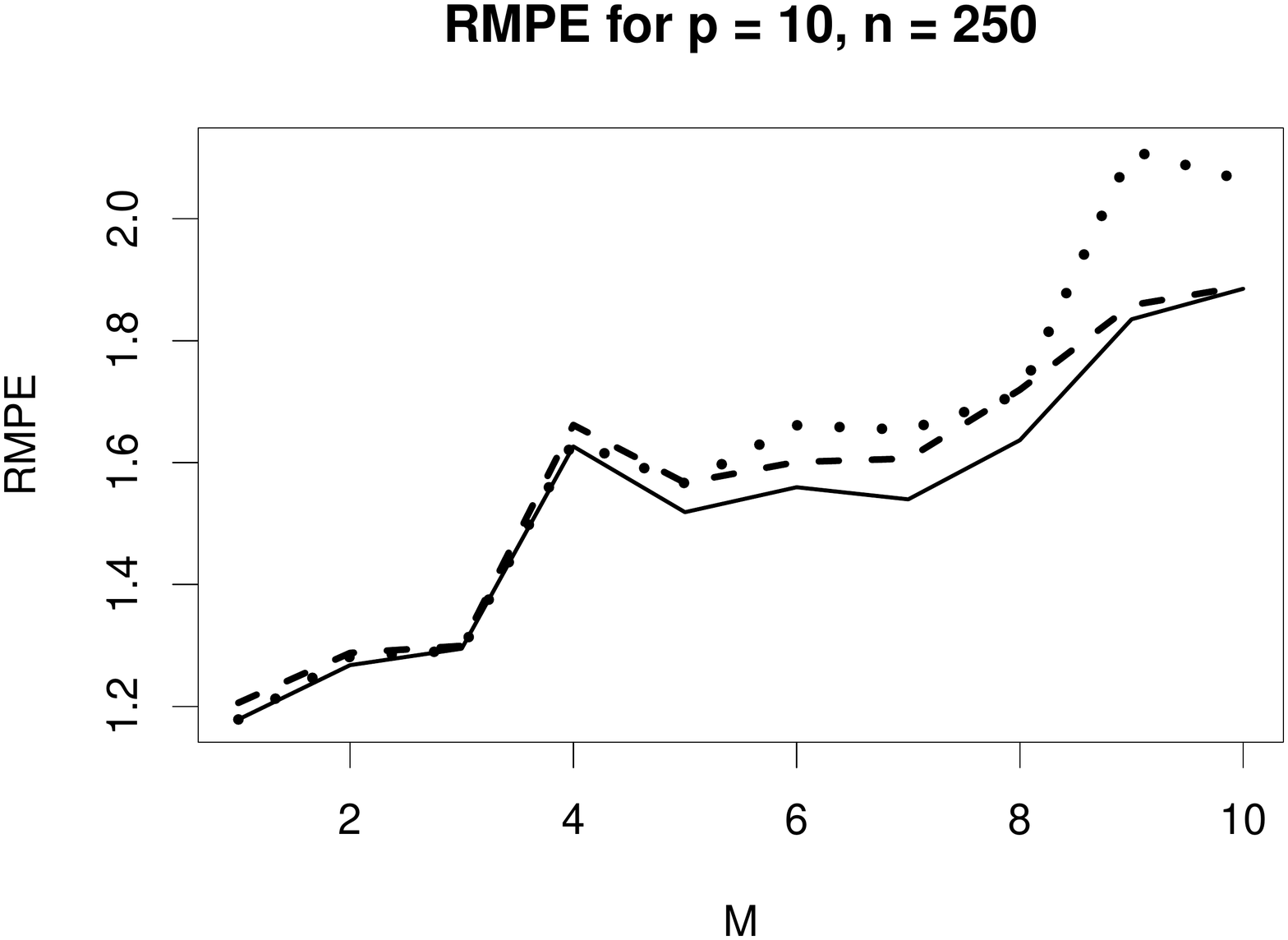}}}
\subfigure[]{\label{fig:rmpesetare}{%
  \includegraphics[height=0.25\textheight]{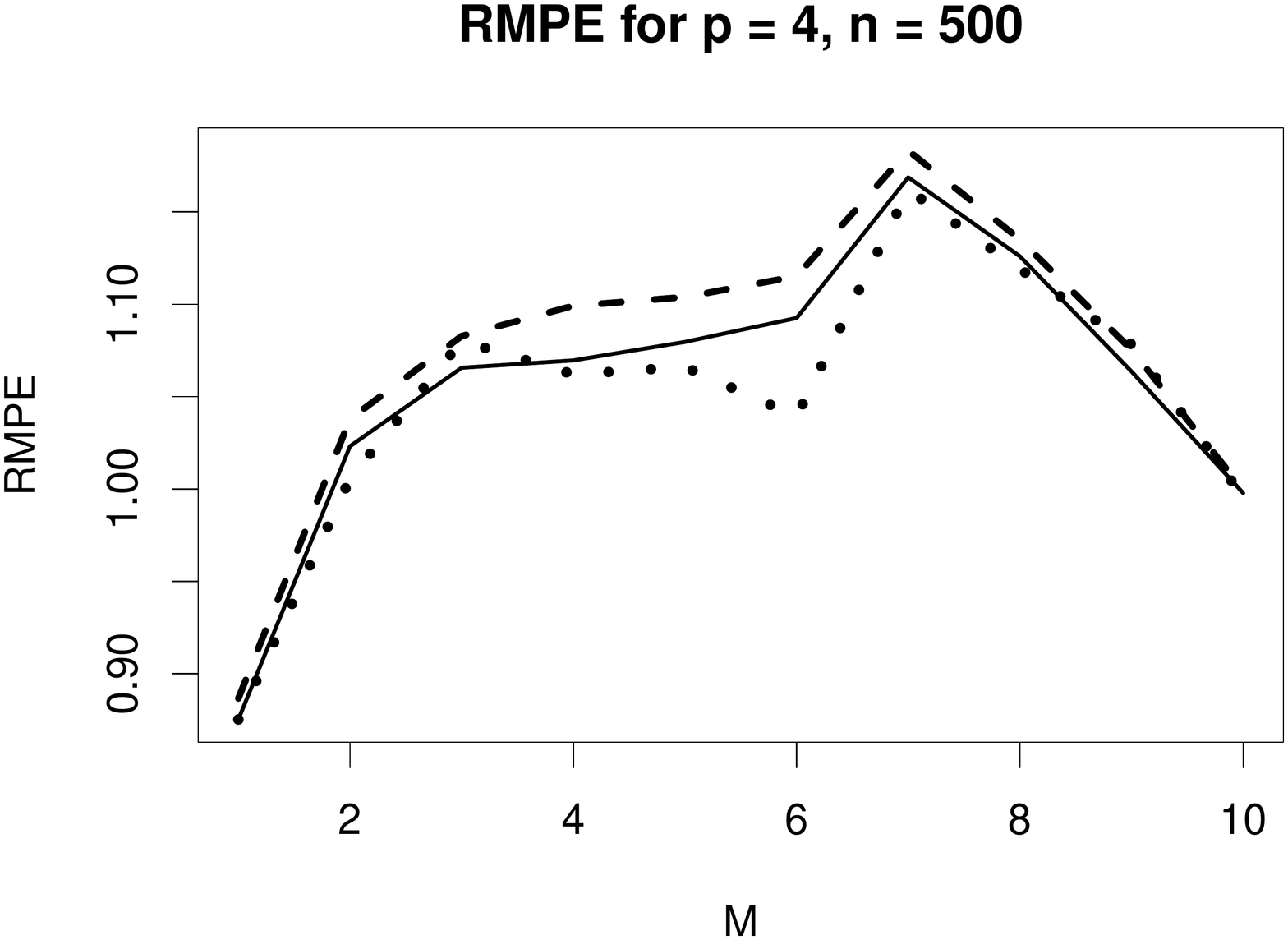}}}
\subfigure[]{\label{fig:rmpesetarf}{%
  \includegraphics[height=0.25\textheight]{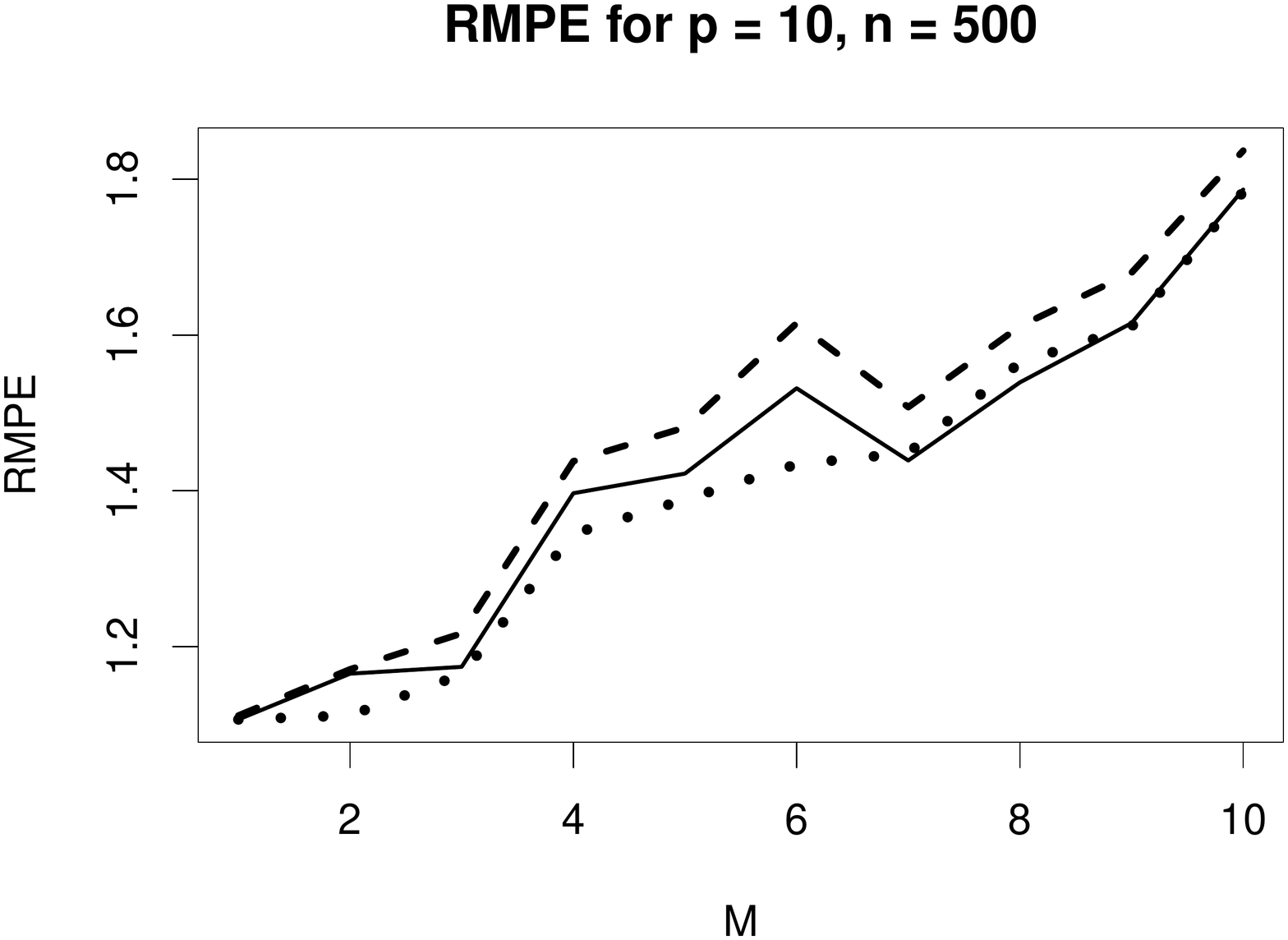}}}
\caption{Plots of the RMPE for Example \ref{ex:setar}.  For each of the plots, the solid line represents the naive forecast, the dotted line represents the bootstrap forecast, and the dashed line represents the multistage forecast. (a) $p=4$, $n=75$, (b) $p=10$, $n=75$, (c) $p=4$, $n=250$, (d) $p=10$, $n=250$, (e) $p=4$, $n=500$, (f) $p=10$, $n=500$.}
\label{fig:rmpesetar}
\end{center}
\end{figure}

\begin{example}
\label{ex:expar}
The next model we consider is the EXPAR model 
\[
X_{t}=\sum_{\alpha=1}^{p}\left(a_{\alpha}+b_{\alpha}\exp\left\{ -\delta X_{t-d}\right\} \right)X_{t-\alpha}+\sigma\left\{ \textbf{X}_{t}\right\} \varepsilon_{t},\qquad\varepsilon_{t}\sim N\left(0,1\right),
\]
where $\sigma\left\{ \textbf{X}_{t}\right\} $ is defined in \eqref{eq:sigma}. We
use 
\[
\textbf{a}=\left(0.3,-0.35,0.1,-0.2\right)^{\prime},
\]
\[
\textbf{b}=\left(0.2,-0.15,0.4,-0.3\right)^{\prime},
\]
and $\delta=25$ for $p=4$, and 
\[
\textbf{a}=\left(0.3,-0.35,0.1,-0.2,0.35,-0.1,.2,-0.3,0.25,-0.25\right)^{\prime},
\]
\[
\textbf{b}=\left(0.2,-0.15,0.4,-0.3,0.15,-0.4,0.3,-0.2,0.25,-0.25\right)^{\prime},
\]
 and $\delta=5$ for $p=10$. For both cases, we set the delay variable
to $X_{t-2}$. As was the case in Example \ref{ex:sine}, the values of $\textbf{a}$
and $\textbf{b}$ are determined so that the bounds of the coefficient
functions ensure the process is geometrically ergodic.  

The estimated coefficient function using the SBLL method and the oracle estimate for $m_{\alpha}\left(X_{t-2}\right)$ with $p=4,10$ and $n=75,500$ are shown in Figures \ref{fig:estexpara} -- \ref{fig:estexpard}. The densities of the empirical efficiencies are shown in Figures \ref{fig:estexpare} and \ref{fig:estexparf}. For this example, the efficiencies are shown for $\alpha=4$ and $\alpha=10$ for $p=4$ and $p=10$, respectively.
As we see in Example \ref{ex:sine}, the mode of the densities tend to be closer to one as $n$ increases.
%
%
%

\begin{center}
{\large{Figure~\ref{fig:estexpar} about here.}}
\end{center}

The RMPEs for the three forecasting methods are shown in Figure \ref{fig:rmpeexpar}. For this example, we see the multistage method having the lowest RMPE for $p=4$ for most values of $M$ (Figures \ref{fig:rmpeexpara}, \ref{fig:rmpeexparc} and \ref{fig:rmpeexpare}). Note the difference between the multistage and the naive methods are small, particularly for $n=500$ (Figure \ref{fig:rmpeexpare}). For $p=10$, we see the naive method have the smallest RMPE for most values of $M$ except for the series length $n=500$ (Figure \ref{fig:rmpeexparf}). For $n=500$, the multistage method tends to have lower RMPE. Again, the differences between the methods are small. 

\begin{center}
{\large{Figure~\ref{fig:rmpeexpar} about here.}}
\end{center}

\begin{example}
\label{ex:setar}
For the last example, we relax the continuity Assumption (A1) and apply the estimation and forecasting methods to the self-exciting threshold autoregressive model (SETAR)
\[
X_{t}=\sum_{\alpha=1}^{p}\phi_{\alpha}\left(X_{t-d}\right)X_{t-\alpha}+\varepsilon_{t},\qquad\varepsilon_{t}\sim N\left(0,1\right),
\]
where 
\[
\phi_{\alpha}\left(X_{t-d}\right)=\begin{cases}
a_{\alpha} & \text{ if }X_{t-d}<r_{\alpha}\\
b_{\alpha} & \text{ if }X_{t-d}\ge r_{\alpha}
\end{cases}.
\]
For $p=4$, we used 
\[
\textbf{a}=\left(0.5,0.2,0.1,-0.4\right)^{\prime},
\]
\[
\textbf{b}=\left(0.4,-0.5,0.5,-0.5\right)^{\prime},
\]
and
\[
\textbf{r}=\left(0,-0.1,-0.2,0\right)^{\prime}.
\]
 For $p=10$, we used 
\[
\textbf{a}=\left(0.5,0.2,0.1,-0.4,0.4,-0.1,-0.2,-0.5,-0.25,0.25\right)^{\prime},
\]
\[
\textbf{b}=\left(0.4,-0.5,0.5,-0.5,0.5,-0.5,0.5,-0.4,0.5,-0.5\right)^{\prime},
\]
and
\[
\textbf{r}=\left(0,-0.1,-0.2,0.1,0.2,0.3,-0.3,0,0.1,0.2\right)^{\prime}.
\]
 For both cases, we used the delay variable $X_{t-1}$. We chose the
values for $\textbf{a}$, $\textbf{b}$, and $\textbf{r}$ to ensure
the process is geometrically ergodic as was we did in Examples \ref{ex:sine} and \ref{ex:expar}.

The estimated coefficient function using the SBLL method and the oracle estimate for $m_{\alpha}\left(X_{t-1}\right)$ with $p=4,10$ and $n=75,500$ are shown in Figures \ref{fig:estsetara} -- \ref{fig:estsetard}. The densities of the empirical efficiencies are shown in Figures \ref{fig:estsetare} and \ref{fig:estsetarf}. For this example, the efficiencies are shown for $\alpha=3$ for both $p=4$ and $p=10$. The estimated coefficient functions looks to be biased at the value of the regime $r_2=-0.1$. 
%
%
%
However, the mode of the efficiencies still tend to be closer to 1 as the series lengths increase indicating that relaxing Assumption (A1) does not affect the asymptotic behavior of the SBLL method, or, at least, it affects the behavior in the same manner as it affects the behavior of the oracle estimator.


\begin{center}
{\large{Figure~\ref{fig:estsetar} about here.}}
\end{center}

The RMPEs for the three forecasting methods are shown in Figure \ref{fig:rmpesetar}. From Figures \ref{fig:rmpesetara} and \ref{fig:rmpesetarb}, we see that the multistage method tends to have the highest RMPE for the larger values of $M$ when the series lengths are $n=75$. This inflated RMPE indicates that the SBLL method is not updating the coefficient functions well in the multistage method. As the series lengths increase (Figures \ref{fig:rmpesetarc} -- \ref{fig:rmpesetarf}), the multistage method becomes less inflated in RMPE. For series length $n=500$, the multistage method has the lowest RMPE for most values of $M$, although, the naive method is close just as it was for the previous two examples.

\begin{center}
{\large{Figure~\ref{fig:rmpesetar} about here.}}
\end{center}

\end{example}

\section{Application to Solar Irradiance}
\label{sec:application}
We now apply the SBLL method to solar irradiance data taken from a sensor located in Ashland, OR, as part of the University of Oregon Solar Radiation Monitoring Laboratory. Many variables affect solar irradiance.  However, one variable that is more influential than others is the amount of cloud cover.  At a typical site, a day (or a period
%
%
of time during a day) is classified as "clear sky," "partly cloudy", or "overcast." In the solar energy literature, methodology used for fitting this type of time series includes ARMIA models \citep{martin2010}, regression analysis \citep{reikard2009}, $k$-nearest neighbors \citep{paoli2010}, and Bayesian models \citep{paoli2010}.  More recently, in \cite{patrick2015}, an FCAR model is used to fit solar irradiance, and is shown to be superior to existing methods in the solar energy literature. 

\begin{figure}[ht]
\begin{center}
\subfigure[]{\label{fig:appa}{%
  \includegraphics[height=0.25\textheight]{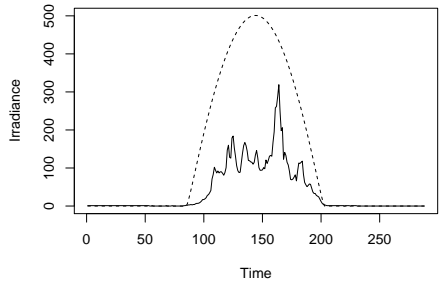}}}
\subfigure[]{\label{fig:appb}{%
  \includegraphics[height=0.25\textheight]{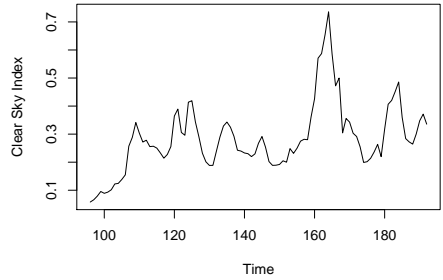}}}
\subfigure[]{\label{fig:appc}{%
  \includegraphics[height=0.25\textheight]{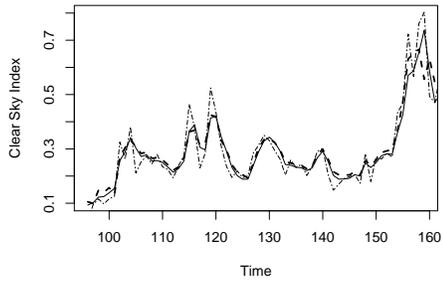}}}
\subfigure[]{\label{fig:appd}{%
  \includegraphics[height=0.25\textheight]{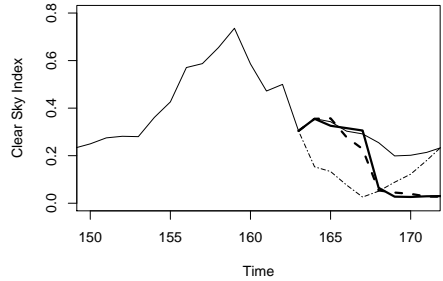}}}
\caption{Plots of (a) measured irradiance (solid line) and clear sky model (dotted line) for November 11, 2013, (b) the clear sky irradiance transformation between 8:00 AM to 4:00 PM, (c) the observed transformed data (solid line) fitted FCAR model using the SBK method (dashed line) and of a linear AR(4) model (dot-dash line), (d) observed transformed data (thin solid line) forecasts for $M=1,\ldots,10$ starting at 1:30 PM using the naive method (thick solid line), the multistage method (dashed line), and a linear AR(4) (dot-dash line) model.}
\label{fig:app}
\end{center}
\end{figure}

The data in this paper contains measured irradiance in $\text{W}/\text{m}^2$ at five minute intervals throughout the day.  Figure~\ref{fig:appa} contains a plot of the data (solid line), with a clear sky model (dotted line) superimposed.  In the measured irradiance, a diurnal trend is clearly present.  
In accordance with the methods described in Section~\ref{sec:estimation} and~\ref{sec:forecasting}, we fit an FCAR model to data taken on November 11, 2013 (a mostly cloudy day) using the SBLL method and forecast ten time points beginning at 1:35 PM (all times are PST). 
%
%

We begin by removing the diurnal trend in the measure irradiance.  Figure \ref{fig:appa} contains a plot of the measured irradiance which are affected by the cloud cover and a theoretical clear sky model; that is, the expected measured irradiance if there were no clouds present in the sky. Even though the data we are using is from a mostly cloudy day, we can still see a diurnal trend that must be removed. A clear sky model is used to remove this trend. A discussion of clear sky irradiance models can be found in \cite{reno2012}. For our application, we used the Ineichen clear sky model \citep{ineichen2002} and calculated the clear sky index with is defined as the ratio of the measured irradiance to the clear sky model. Figure \ref{fig:appb} shows the clear sky index for the time interval starting at 8:00 AM ($t=96$) and ending at 4:00 PM ($t=192$). 

\begin{center}
{\large{Figure~\ref{fig:app} about here.}}
\end{center}

%
%

The fit of the FCAR model can be seen in Figure \ref{fig:appc}. We found that $p=2$ and $d=5$ gave the best fit. We also fit a linear AR model of order 4 which we determined by minimizing the AIC. The MSE's of the fitted models were 0.0009 for the FCAR model and 0.0020 for the linear AR model. 

We forecast for $M=1,\ldots,10$ using the naive and multistage methods and compared them to forecasts using the linear AR model. Figure \ref{fig:appd} shows the forecasts and the observed values starting at 1:35 PM ($t=163$). The root squared prediction errors (RSPE) for the three methods are given in Table \ref{tab:table1}. For $M=1,\ldots, 5$, the FCAR forecasting methods had lower RSPEs with the naive method having the lowest at $M=1,3,4,5$ and the multistage method have the lowest at $M=2$. At $M=6,\ldots,10$, the linear AR model had the lowest RSPE. For the smaller values of $M$, these results agree with the simulation results in that the naive method performs just as well if not better than the multistage method.

\begin{center}
{\large{Table~\ref{tab:table1} about here.}}
\end{center}

\section{Discussion}
\label{sec:discussion}
We have adapted the SBLL method to FCAR models and shown that these estimators are oracally efficient. We examined the performance of naive, bootstrap, and multistage forecasting methods with a model estimated with the SBLL method. By estimating the model in this way, we have shown through simulation results that the bootstrap method did not perform as well as the other two methods. We have also shown that the naive method performs just as well as the multistage method and even outperforms it in some situations. The main advantage to using the naive method is that it is much faster computationally than the other two methods. 

For a real world example, we showed the naive and multistage methods performing better than a linear AR model when applied to solar irradiance data for $M=1,\ldots,6$. The day we selected for our application was mostly cloudy throughout the day. Future research will examine the fit of an FCAR model to irradiance data when the cloud cover conditions change during the day. This model will need to incorporate a covariate for the cloud cover. In this paper, we have shown that the SBLL method is adequate for fitting the model and forecasting in the absence of a covariate as long as the amount of cloud cover is constant throughout the day.  

\section*{Acknowledgments}
  
The research was performed under contract (PO 1303122) with Sandia
National Laboratories, a multi-program laboratory managed and operated
by Sandia Corporation, a wholly owned subsidiary of Lockheed Martin
Corporation, for the U.S.~Department of Energy's National Nuclear
Security Administration under contract DE-AC04-94AL85000.

\newpage
\bibliographystyle{model2-names}
\bibliography{references}

\newpage
\begin{figure}[ht]
\begin{center}
\subfigure[]{\label{fig:estsinea}{%
  \includegraphics[height=0.25\textheight]{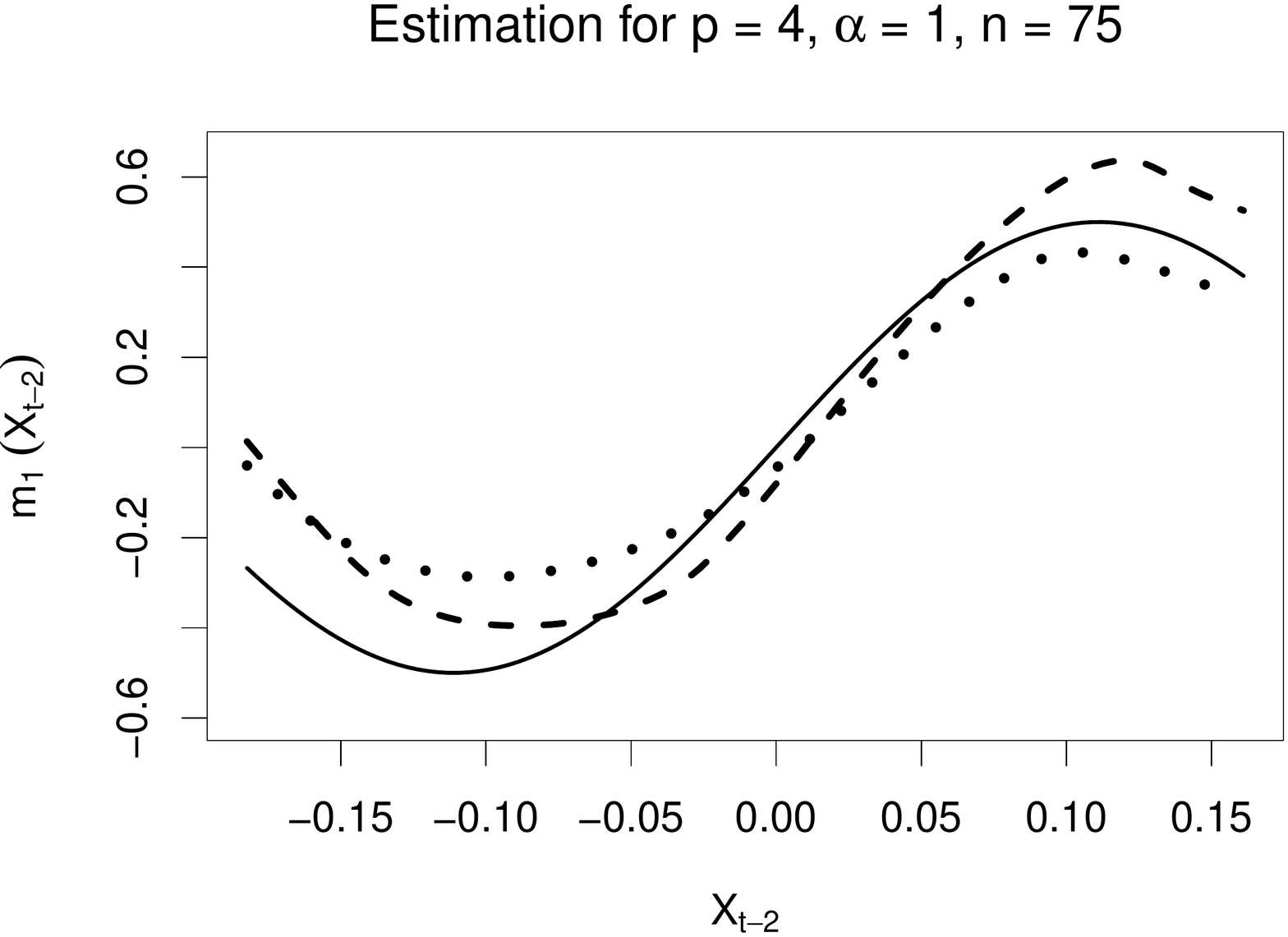}}}
\subfigure[]{\label{fig:estsineb}{%
  \includegraphics[height=0.25\textheight]{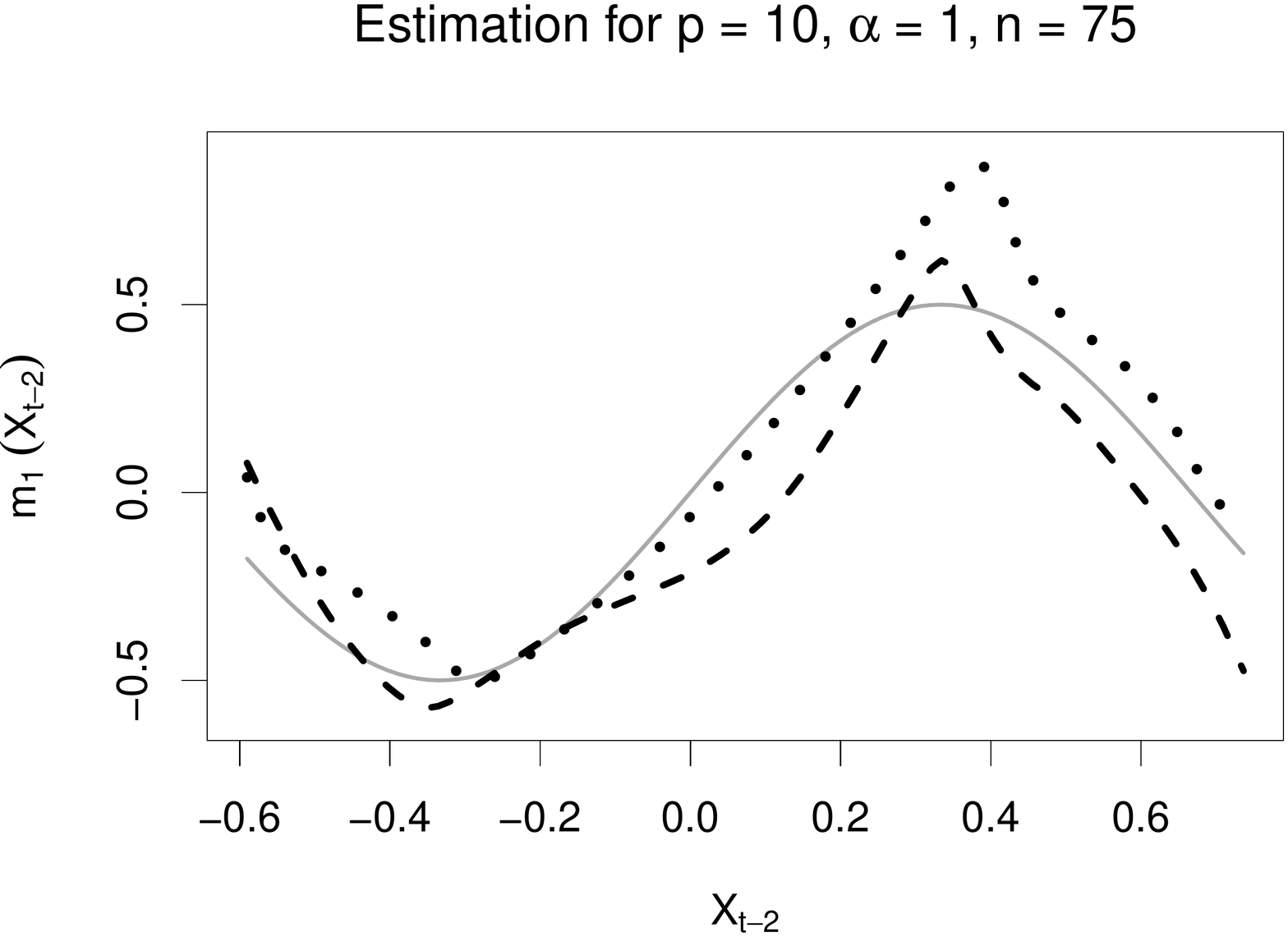}}}
\subfigure[]{\label{fig:estsinec}{%
  \includegraphics[height=0.25\textheight]{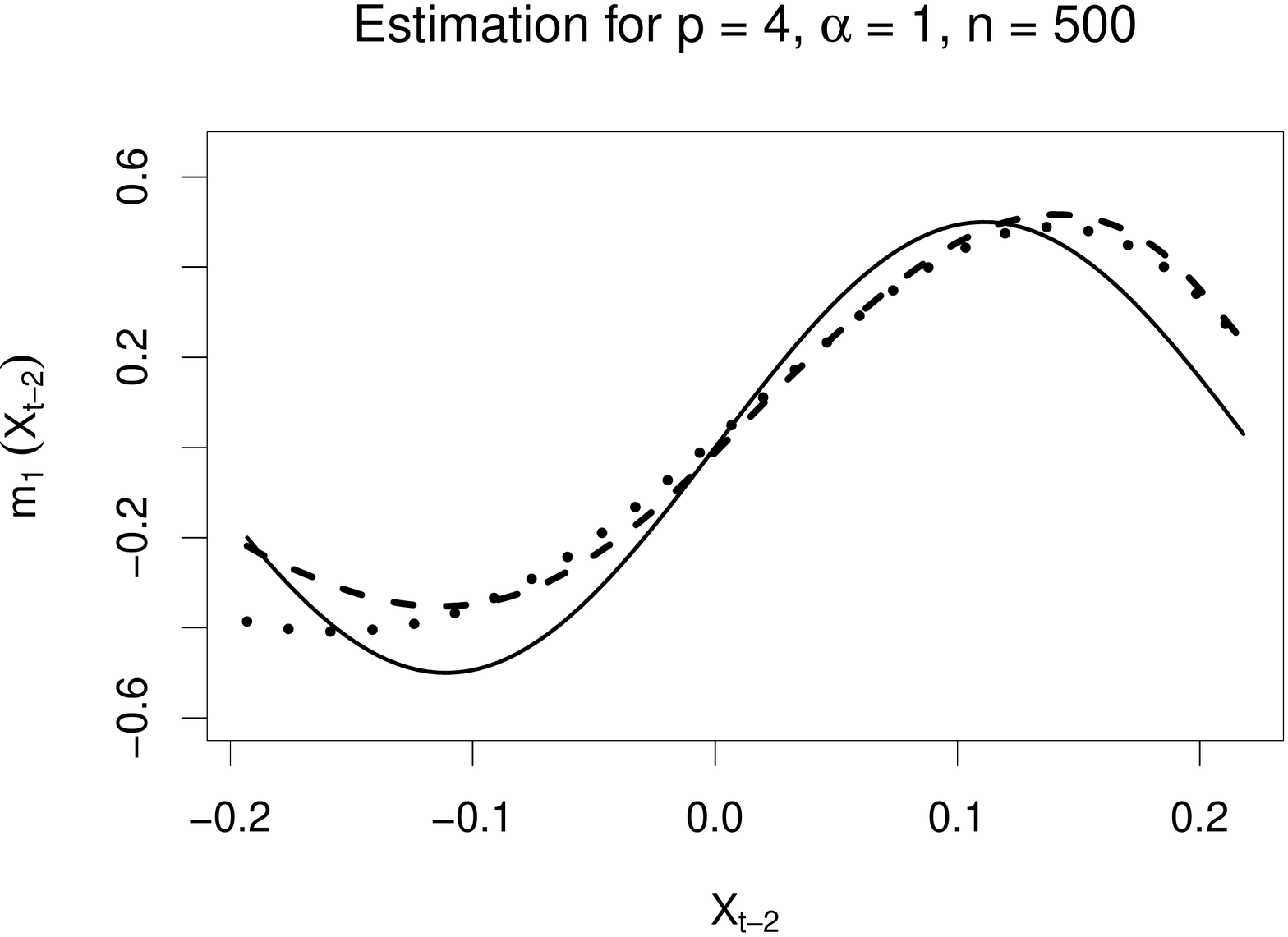}}}
\subfigure[]{\label{fig:estsined}{%
  \includegraphics[height=0.25\textheight]{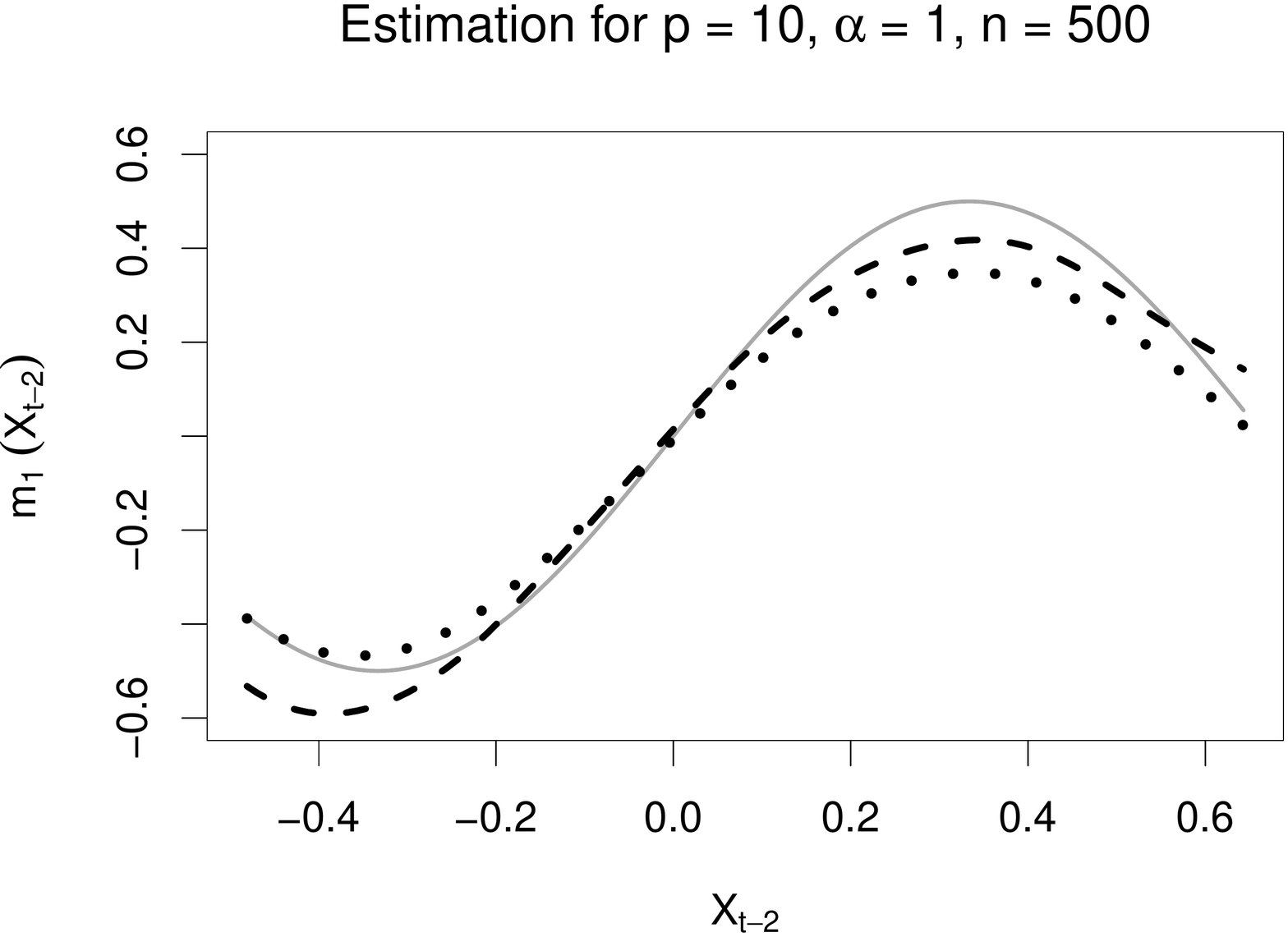}}}
\subfigure[]{\label{fig:estsinee}{%
  \includegraphics[height=0.25\textheight]{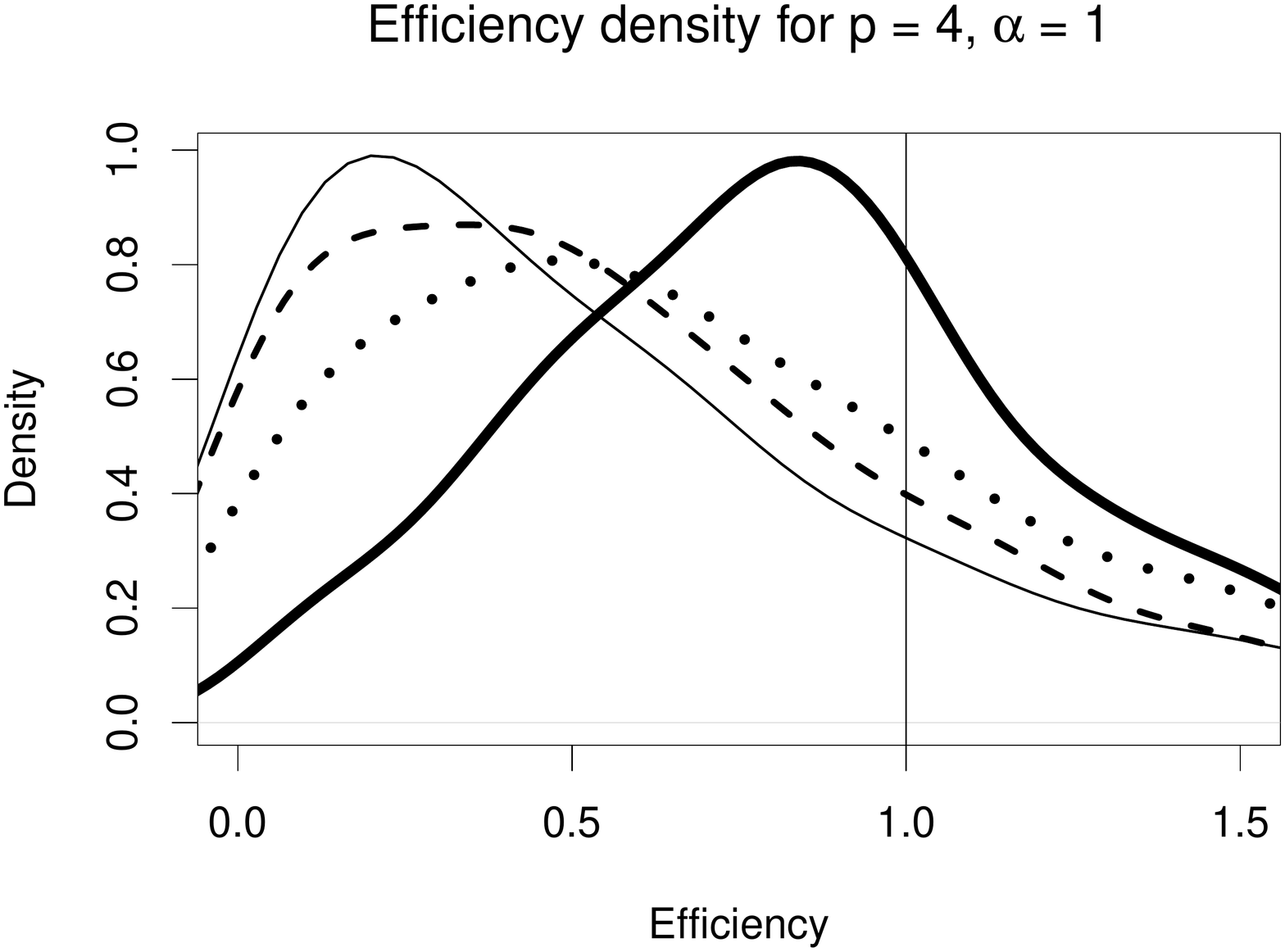}}}
\subfigure[]{\label{fig:estsinef}{%
  \includegraphics[height=0.25\textheight]{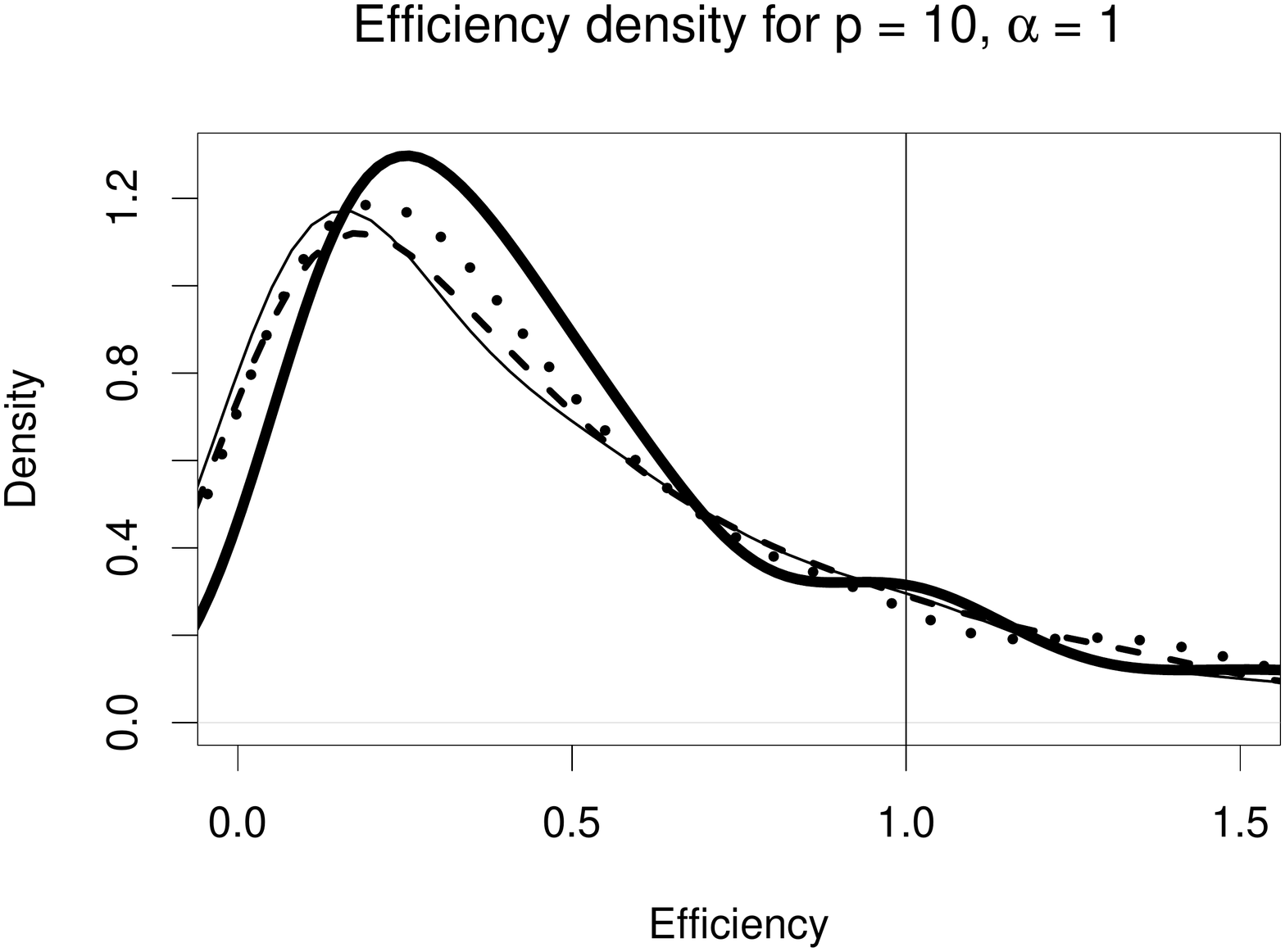}}}
\caption{Plots (a) -- (d) are graphs of the true coefficient function (solid line), the SBLL estimate (dashed line), and the oracle estimate (dotted line) for Example \ref{ex:sine} : (a) $p=4$, $n=75$, (b) $p=10$, $n=75$, (c) $p=4$, $n=500$, (d) $p=10$, $n=500$. Plots (e) and (f) contain the empirical efficiency densities for Example \ref{ex:sine} with series lengths $n=75$ (thin solid line), $150$ (dashed line), $250$ (dotted line), and $500$ (thick solid line) for $p=4$ (e) and $p=10$ (f).}
\label{fig:estsine}
\end{center}
\end{figure}

\newpage
\begin{figure}[ht]
\begin{center}
\subfigure[]{\label{fig:rmpesinea}{%
  \includegraphics[height=0.25\textheight]{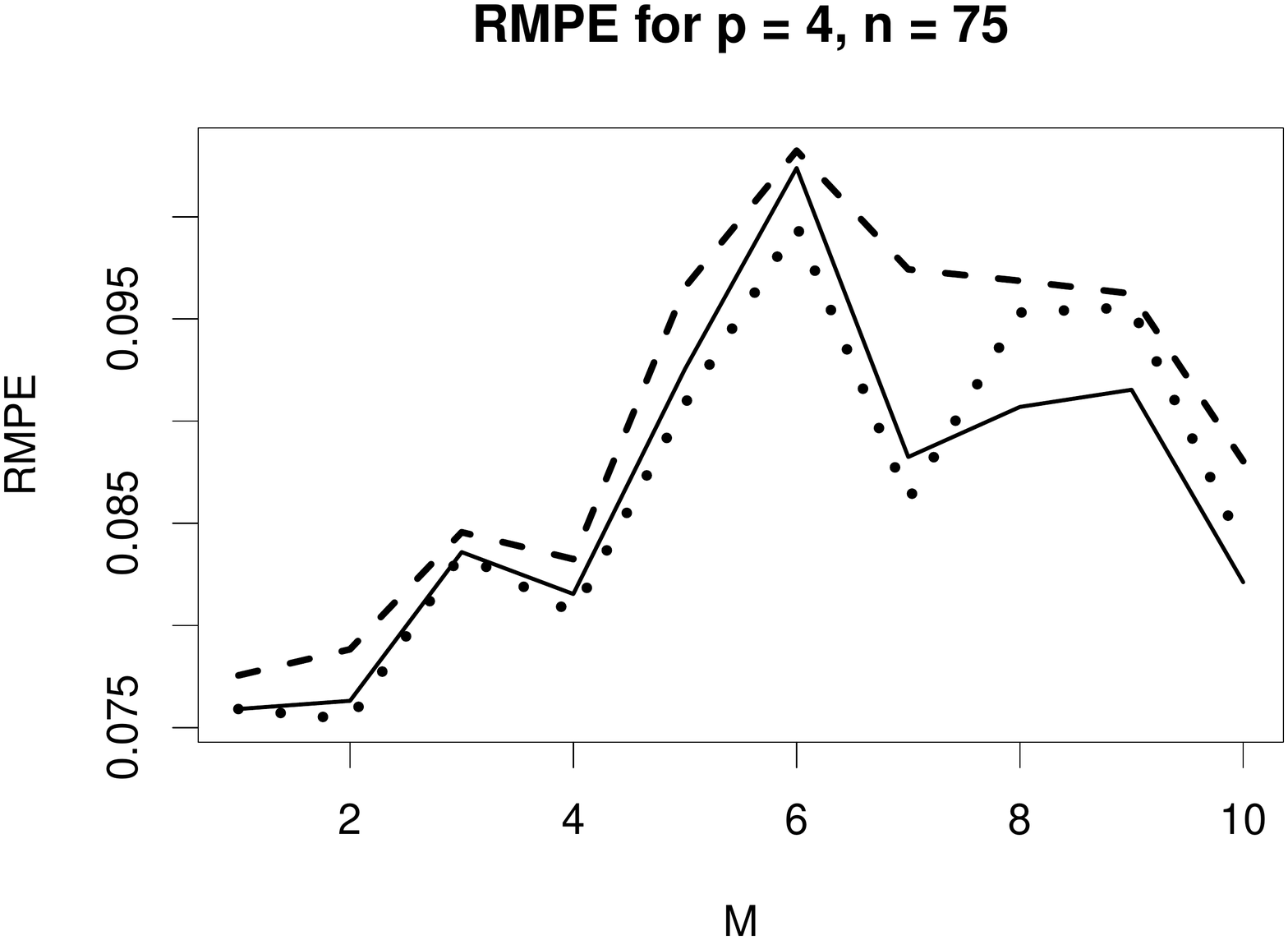}}}
\subfigure[]{\label{fig:rmpesineb}{%
  \includegraphics[height=0.25\textheight]{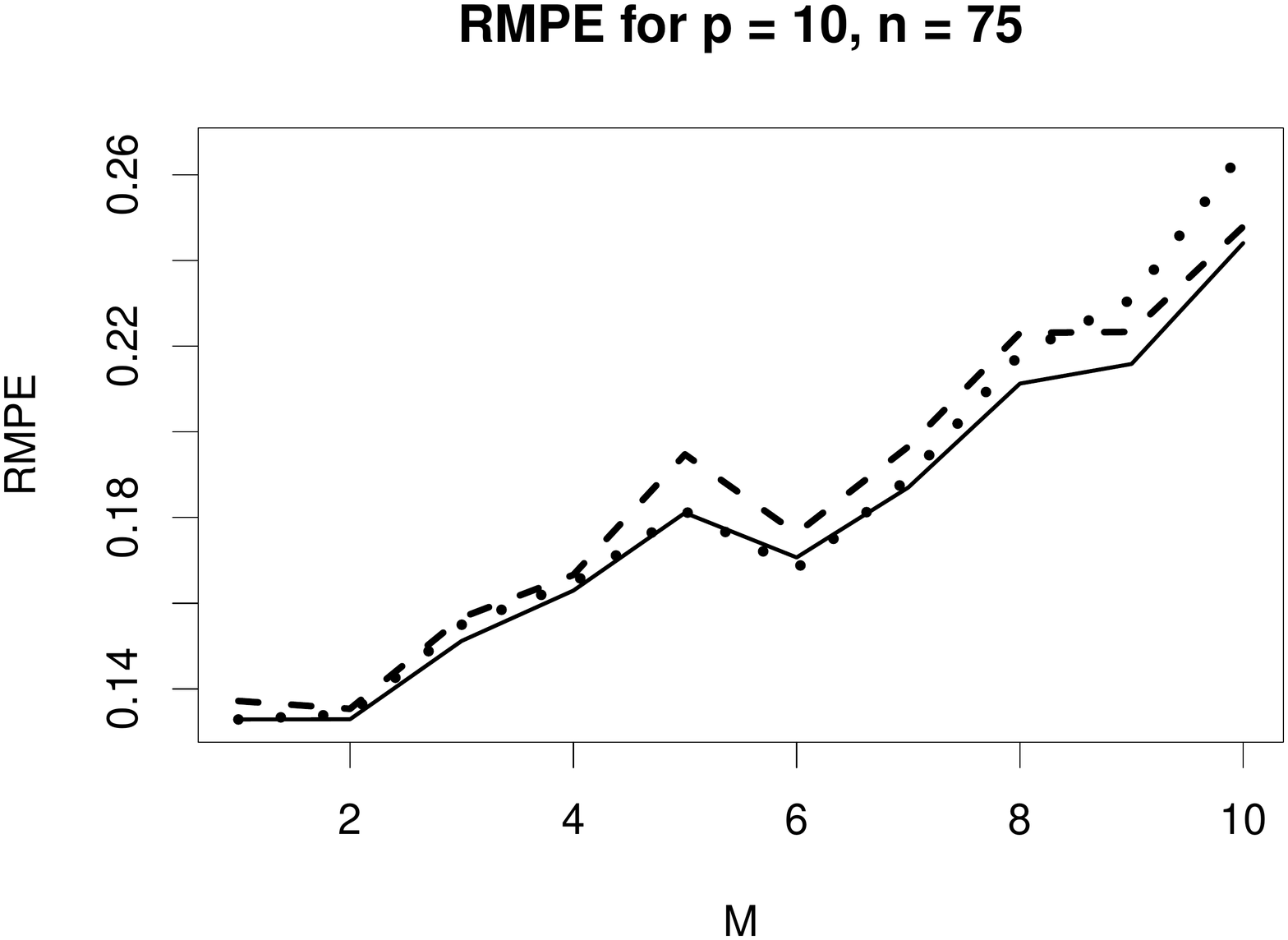}}}
\subfigure[]{\label{fig:rmpesinec}{%
  \includegraphics[height=0.25\textheight]{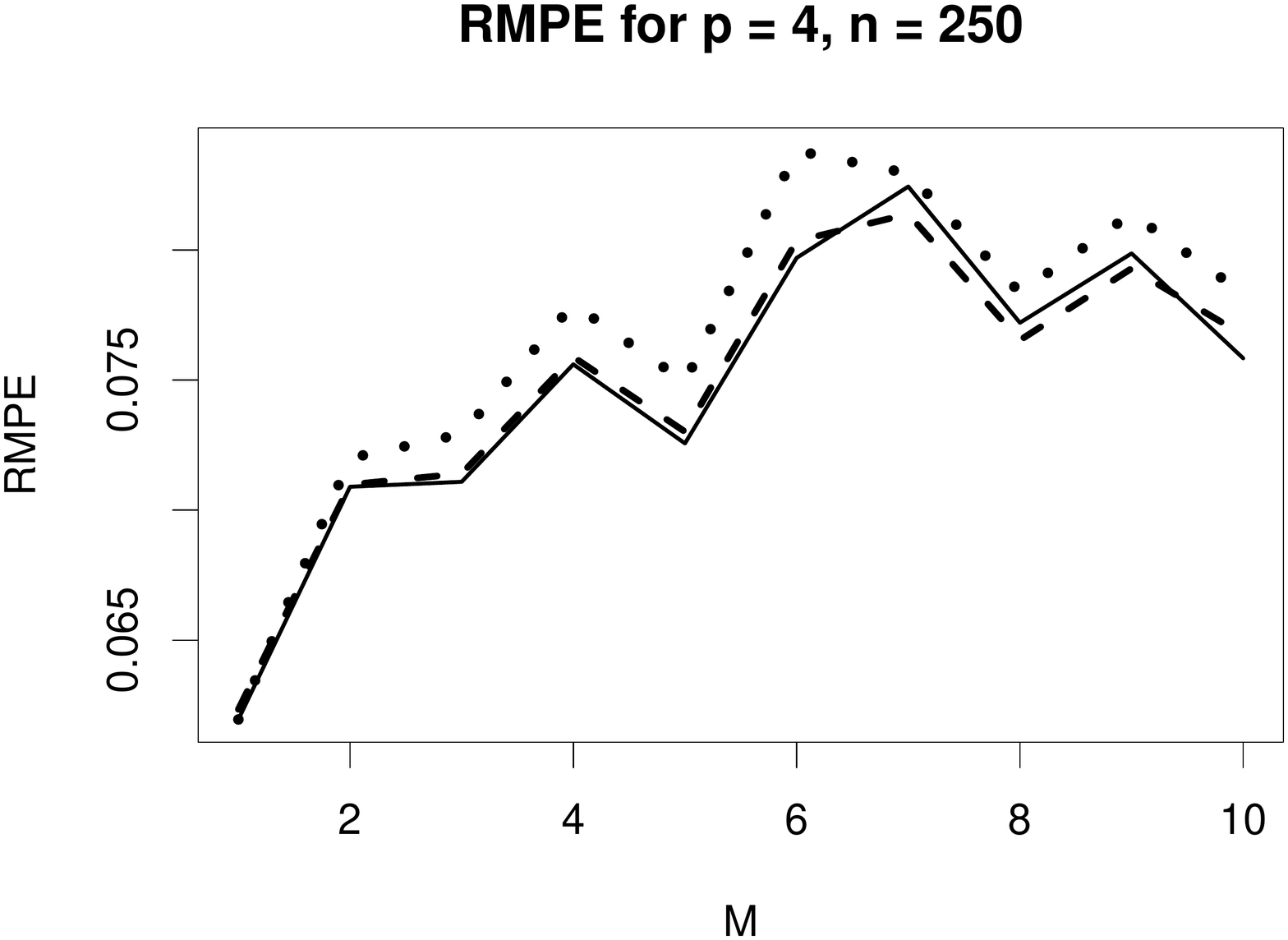}}}
\subfigure[]{\label{fig:rmpesined}{%
  \includegraphics[height=0.25\textheight]{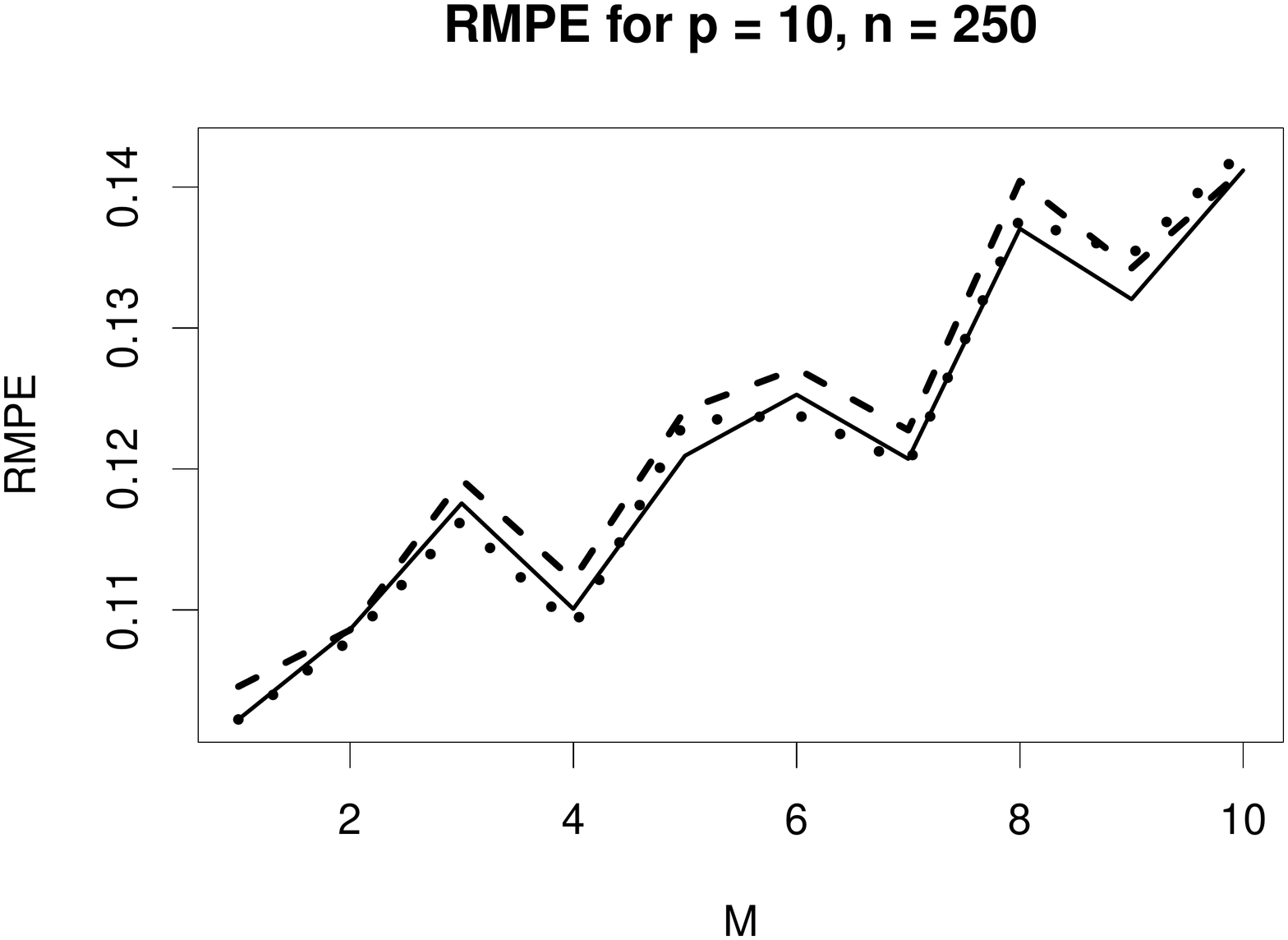}}}
\subfigure[]{\label{fig:rmpesinee}{%
  \includegraphics[height=0.25\textheight]{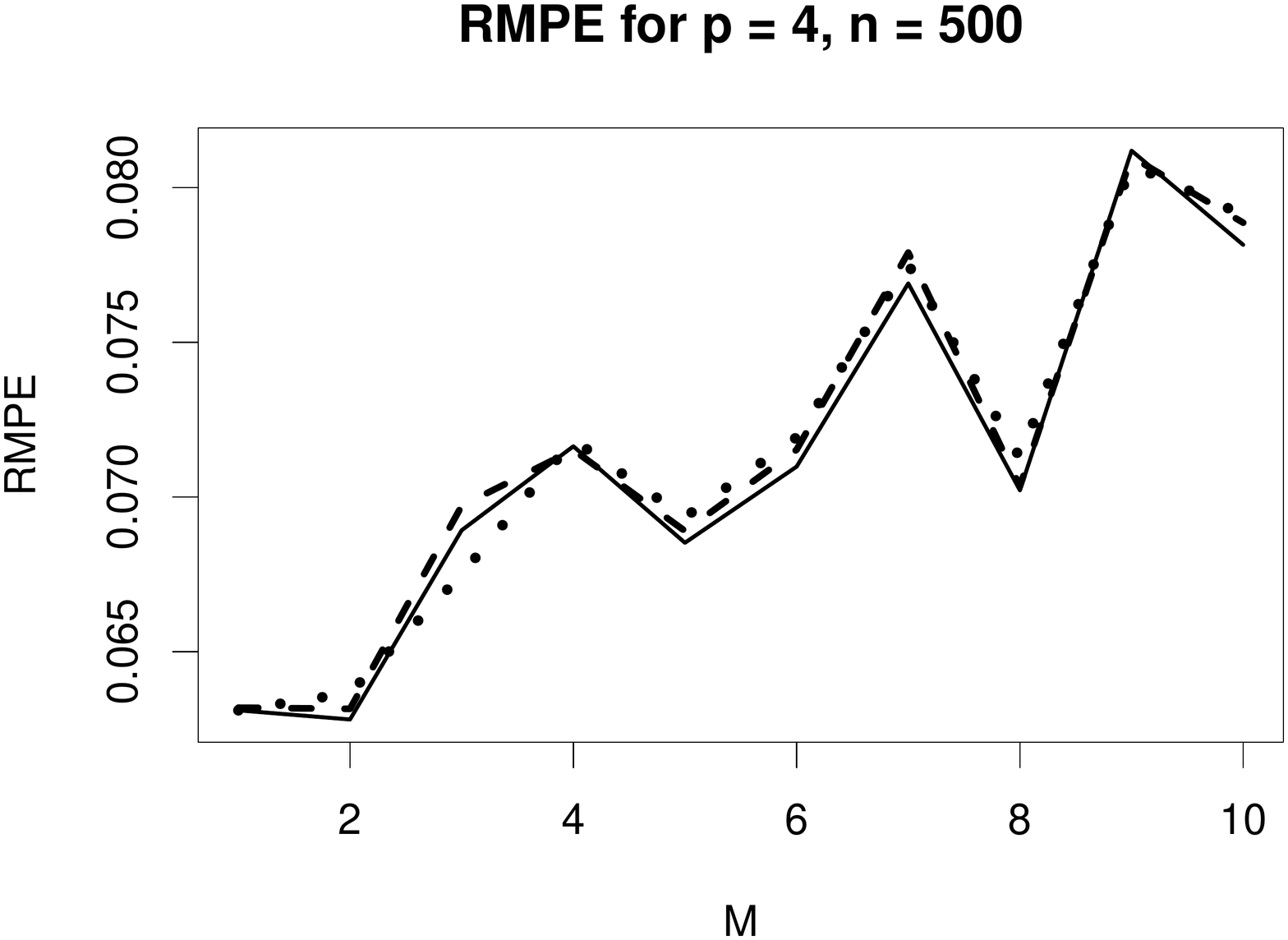}}}
\subfigure[]{\label{fig:rmpesinef}{%
  \includegraphics[height=0.25\textheight]{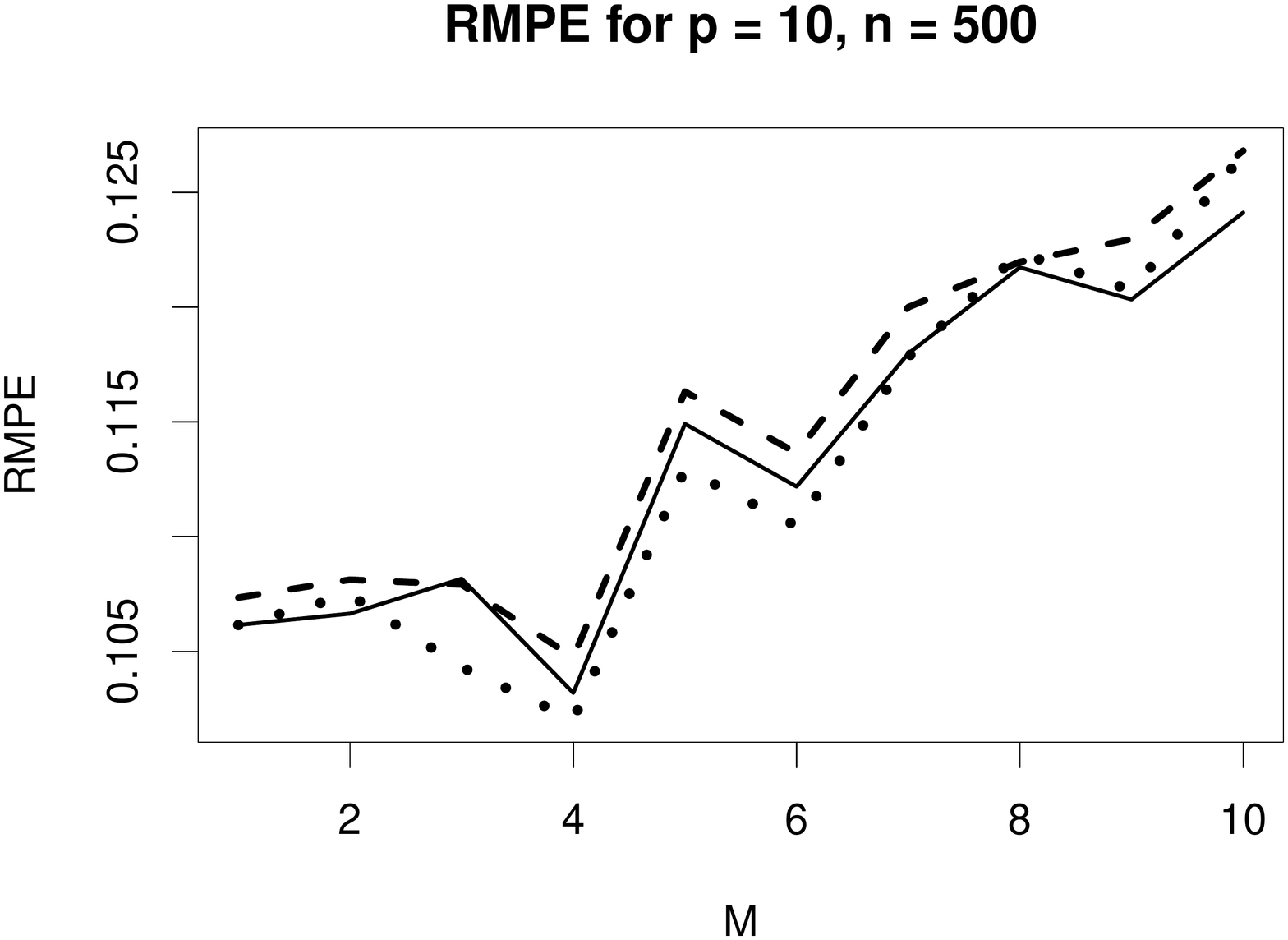}}}
\caption{Plots of the RMPE for Example \ref{ex:sine}.  For each of the plots, the solid line represents the naive forecast, the dashed line represents the bootstrap forecast, and the dotted line represents the multistage forecast.  (a) $p=4$, $n=75$, (b) $p=10$, $n=75$, (c) $p=4$, $n=250$, (d) $p=10$, $n=250$, (e) $p=4$, $n=500$, (f) $p=10$, $n=500$.}
\label{fig:rmpesine}
\end{center}
\end{figure}

\newpage
\begin{figure}[ht]
\begin{center}
\subfigure[]{\label{fig:estexpara}{%
  \includegraphics[height=0.25\textheight]{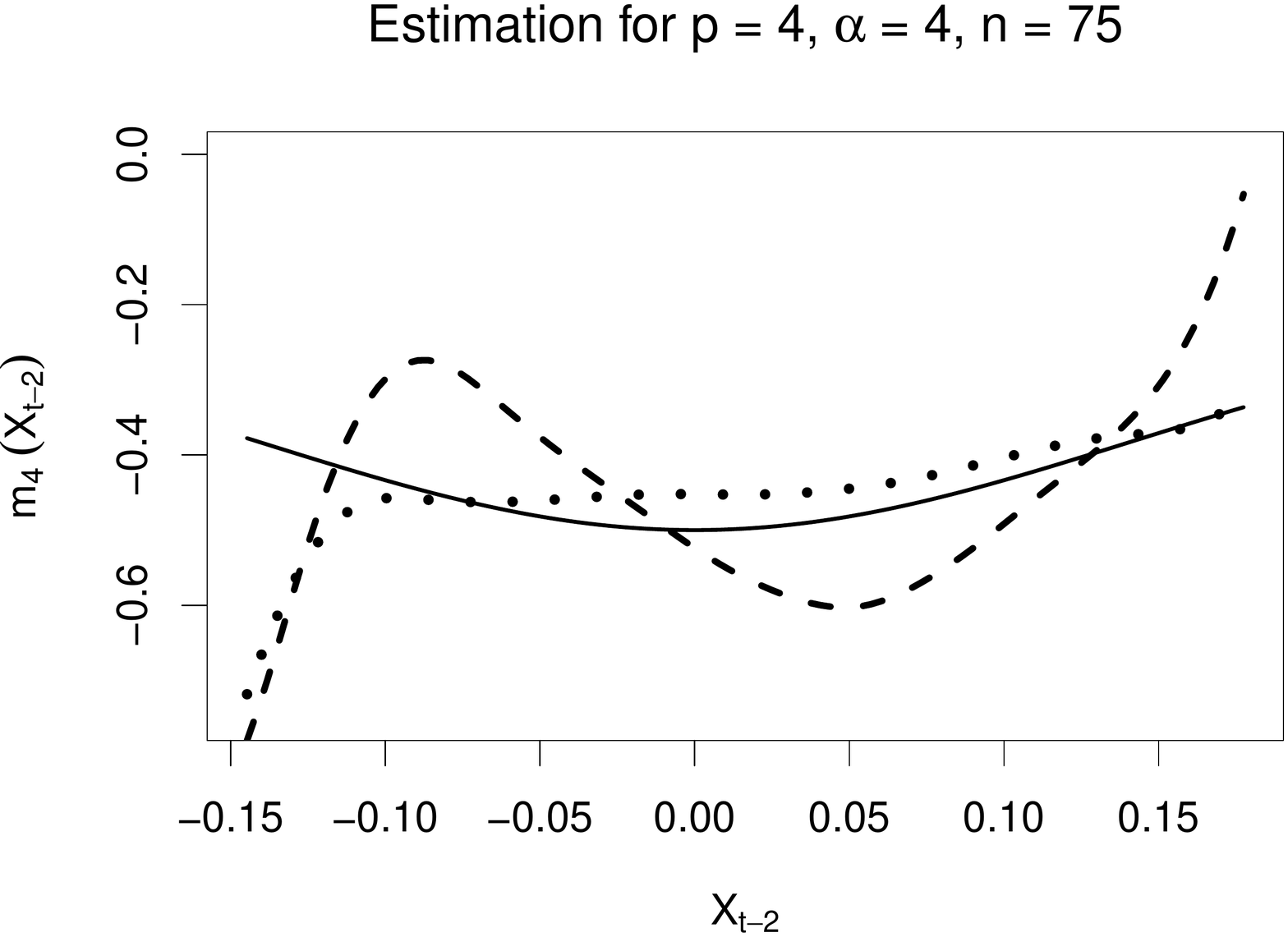}}}
\subfigure[]{\label{fig:estexparb}{%
  \includegraphics[height=0.25\textheight]{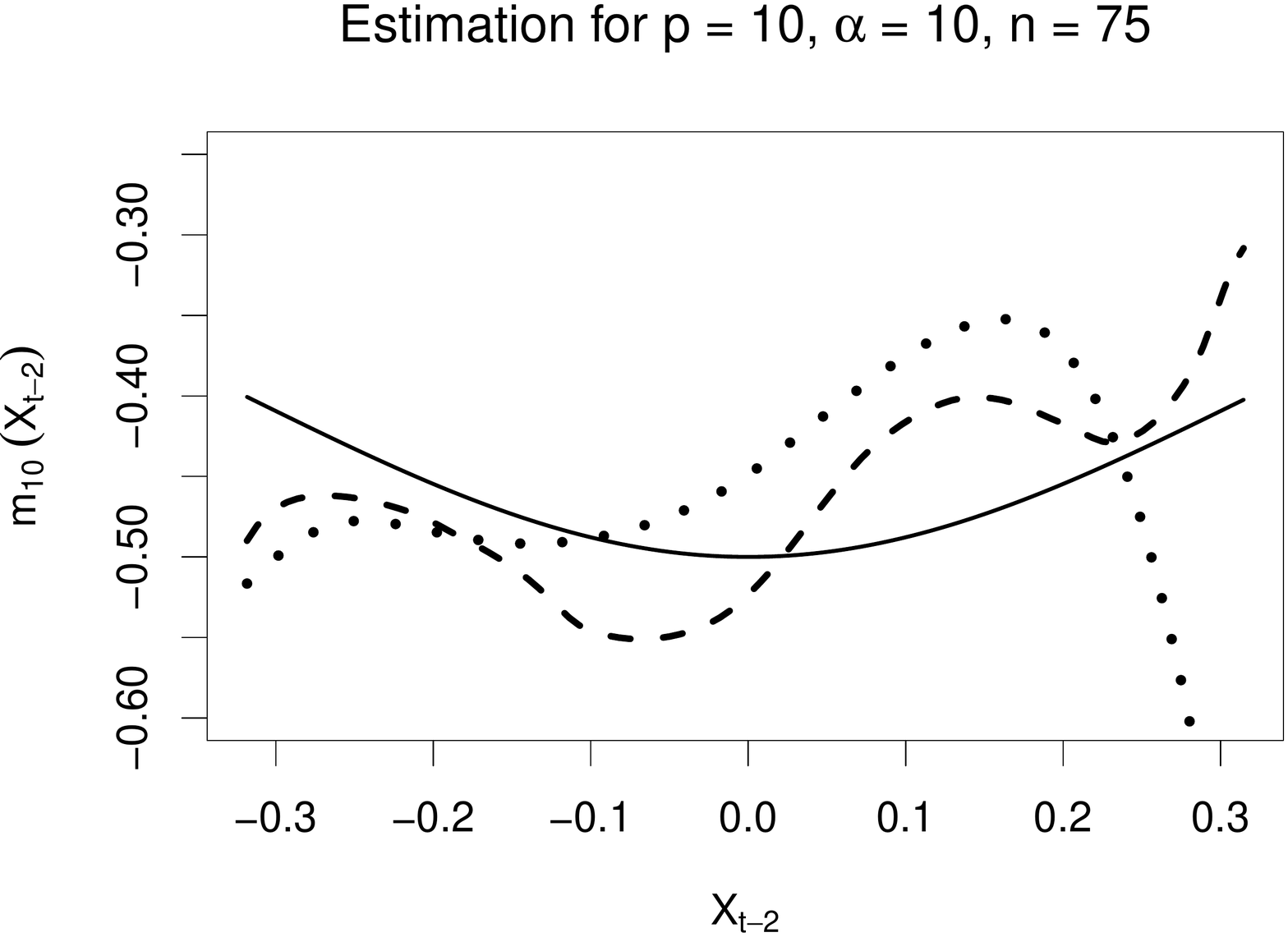}}}
\subfigure[]{\label{fig:estexparc}{%
  \includegraphics[height=0.25\textheight]{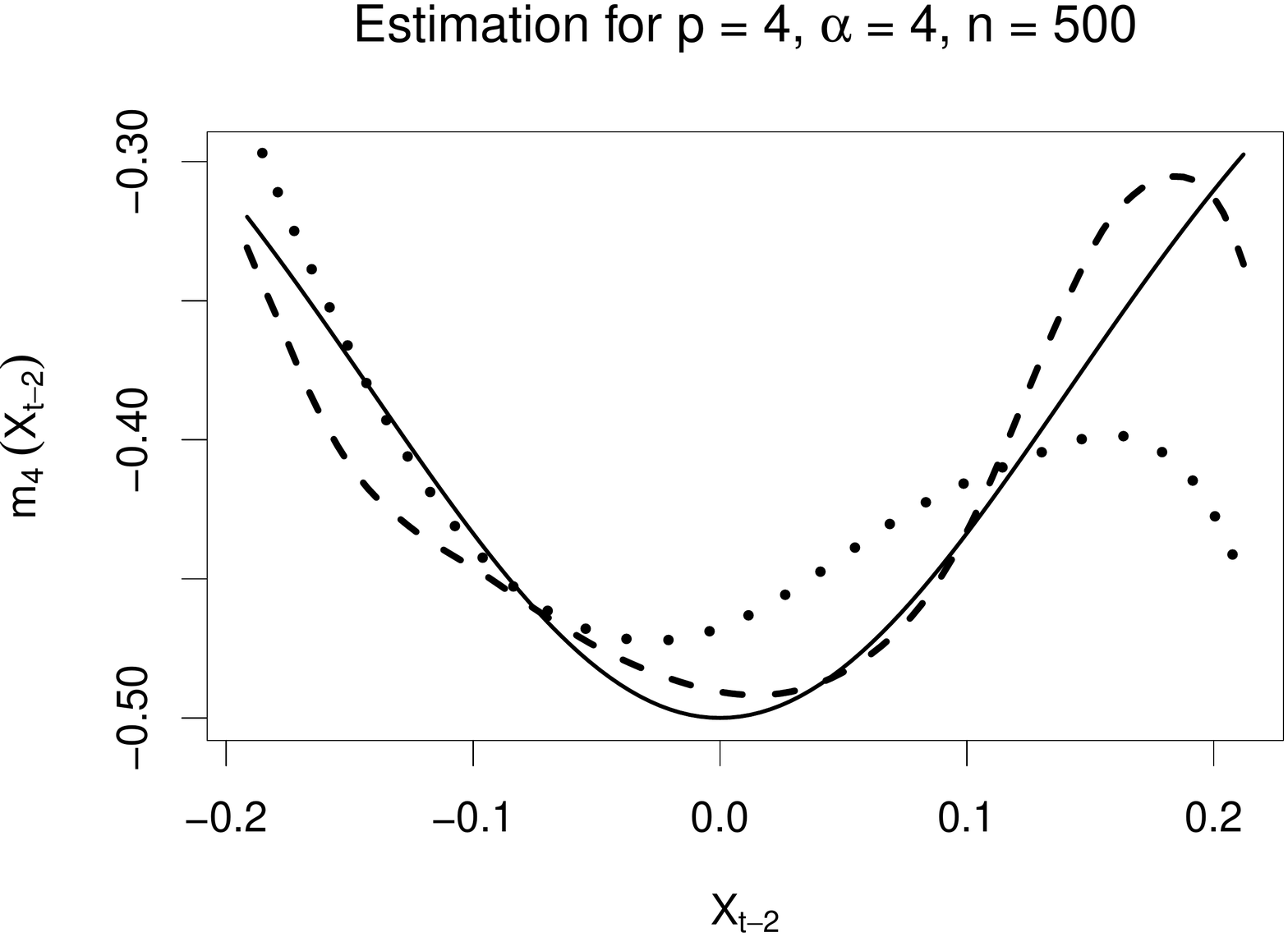}}}
\subfigure[]{\label{fig:estexpard}{%
  \includegraphics[height=0.25\textheight]{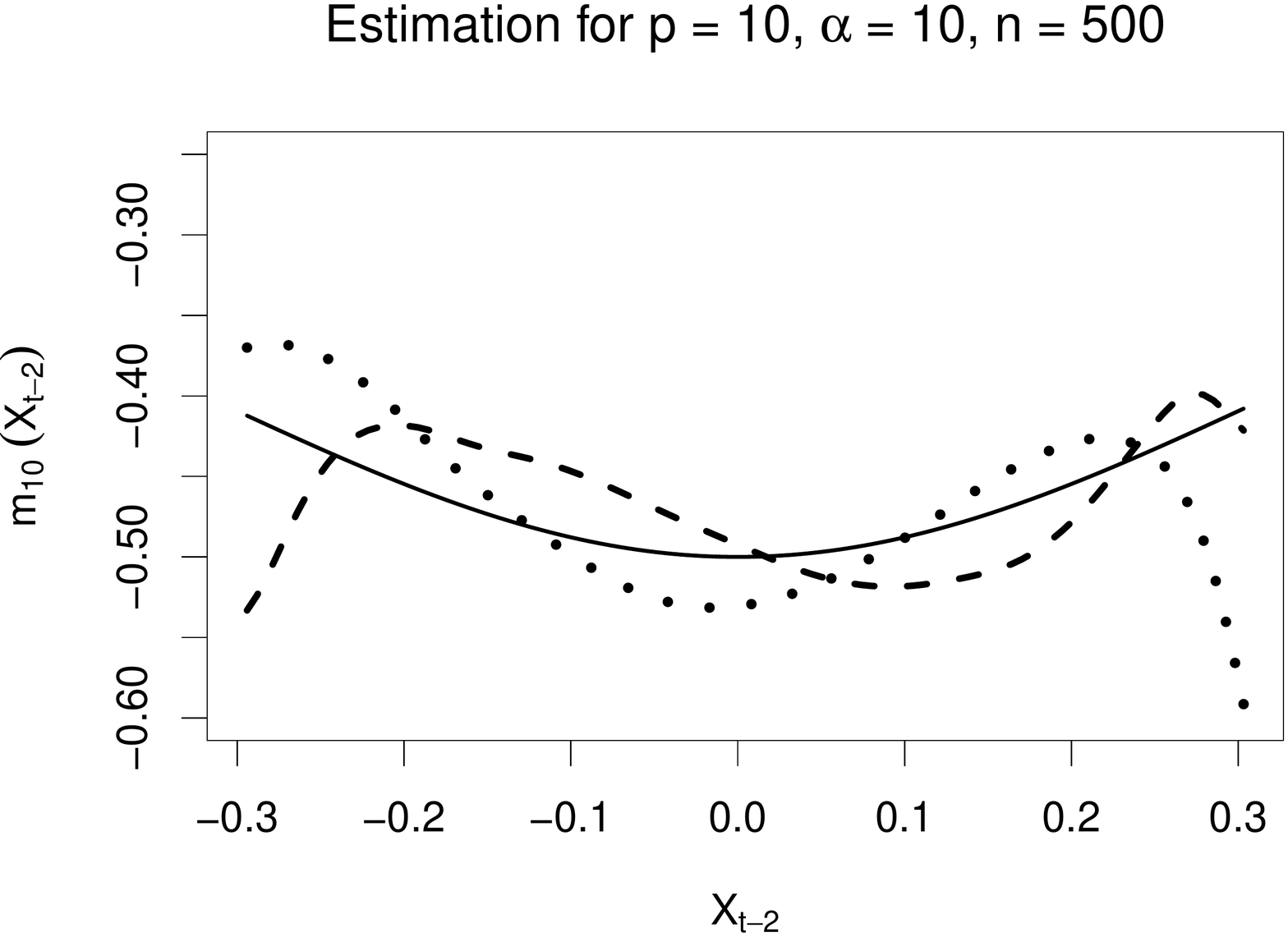}}}
\subfigure[]{\label{fig:estexpare}{%
  \includegraphics[height=0.25\textheight]{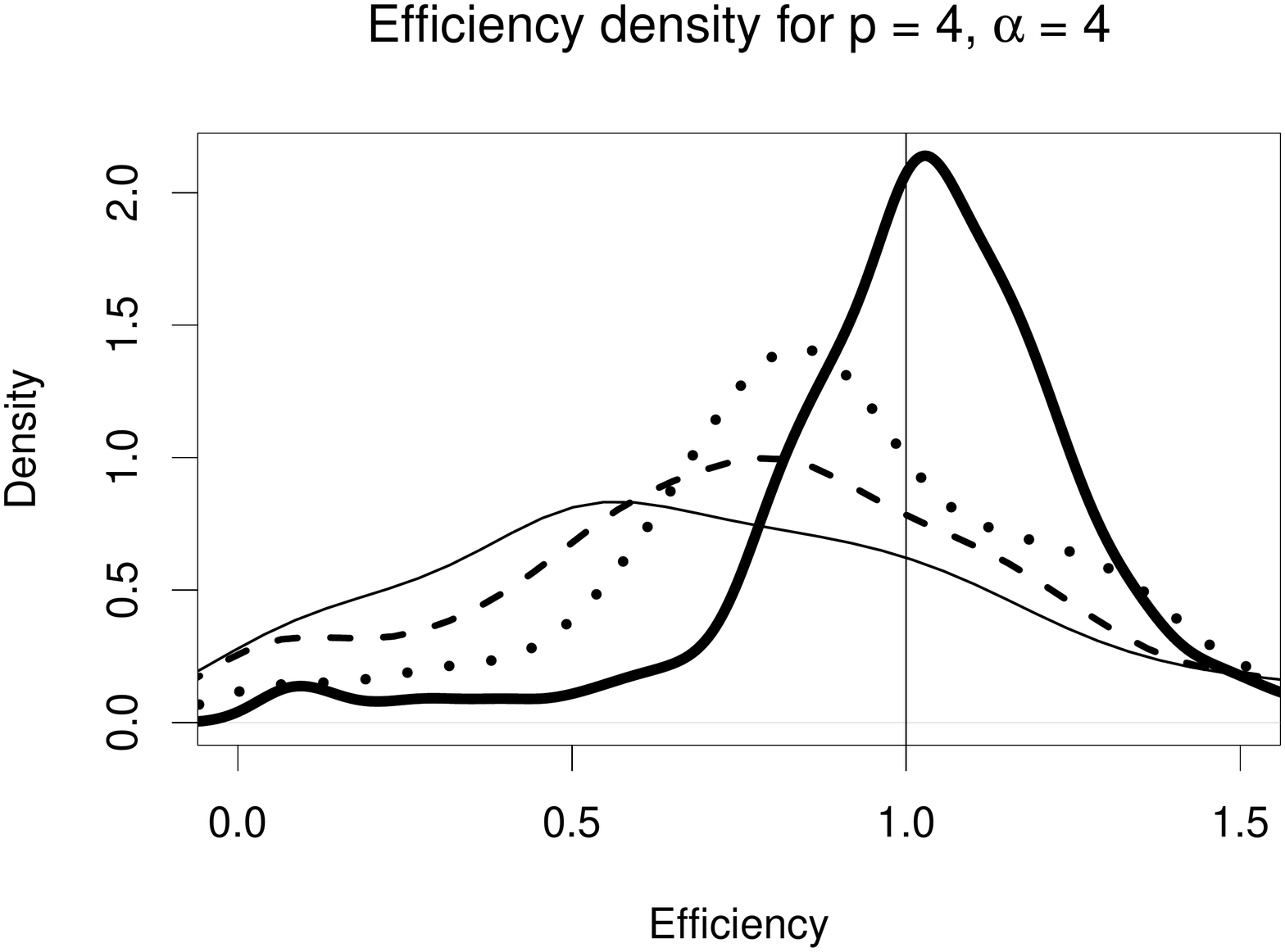}}}
\subfigure[]{\label{fig:estexparf}{%
  \includegraphics[height=0.25\textheight]{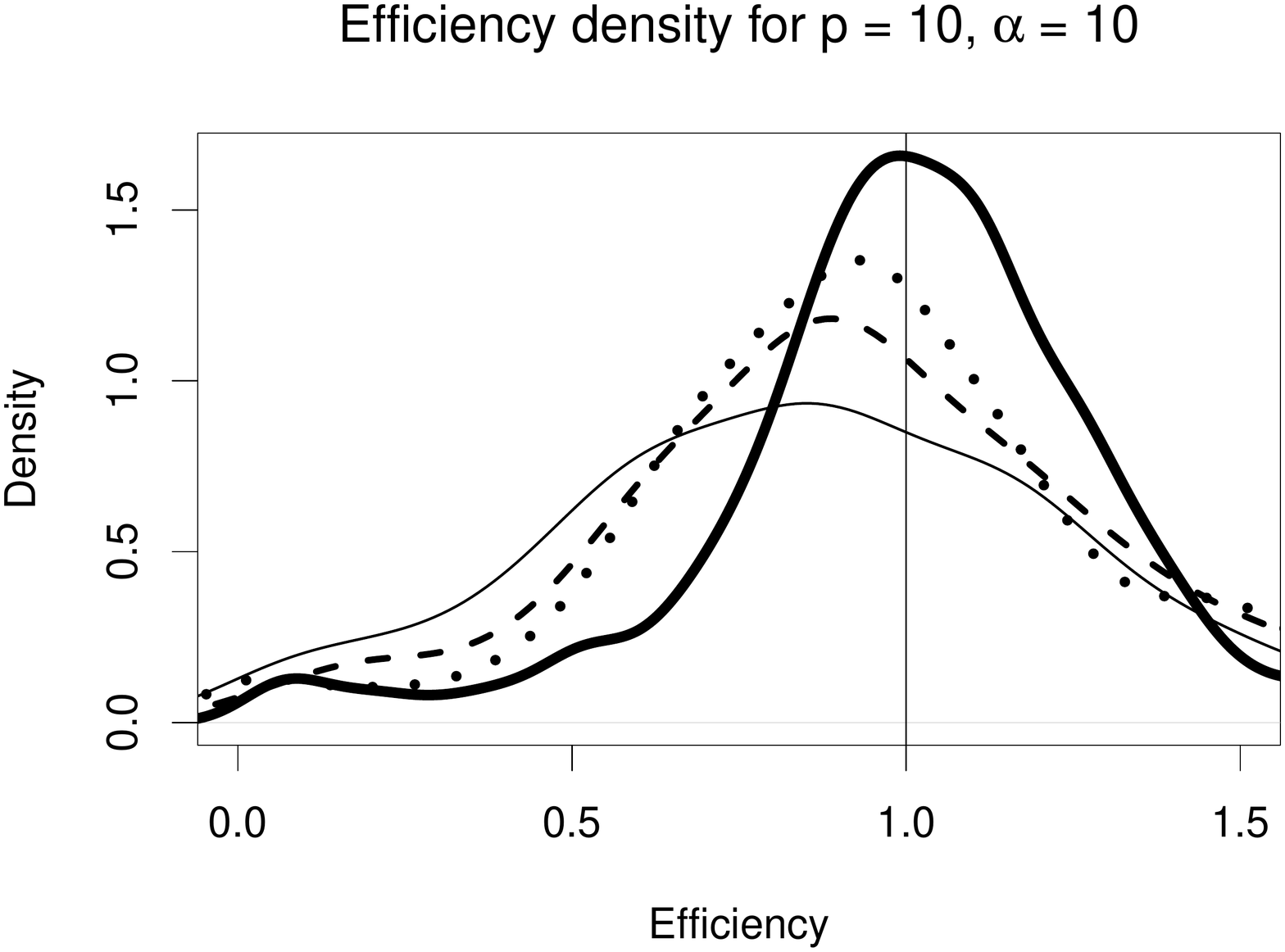}}}
\caption{Plots (a) -- (d) are graphs of the true coefficient function (solid line), the SBLL estimate (dashed line), and the oracle estimate (dotted line) for Example \ref{ex:expar} : (a) $p=4$, $n=75$, (b) $p=10$, $n=75$, (c) $p=4$, $n=500$, (d) $p=10$, $n=500$. Plots (e) and (f) contain the empirical efficiency densities for Example \ref{ex:expar} with series lengths $n=75$ (thin solid line), $150$ (dashed line), $250$ (dotted line), and $500$ (thick solid line) for $p=4$ (e) and $p=10$ (f).}
\label{fig:estexpar}
\end{center}
\end{figure}

\newpage
\begin{figure}[ht]
\begin{center}
\subfigure[]{\label{fig:rmpeexpara}{%
  \includegraphics[height=0.25\textheight]{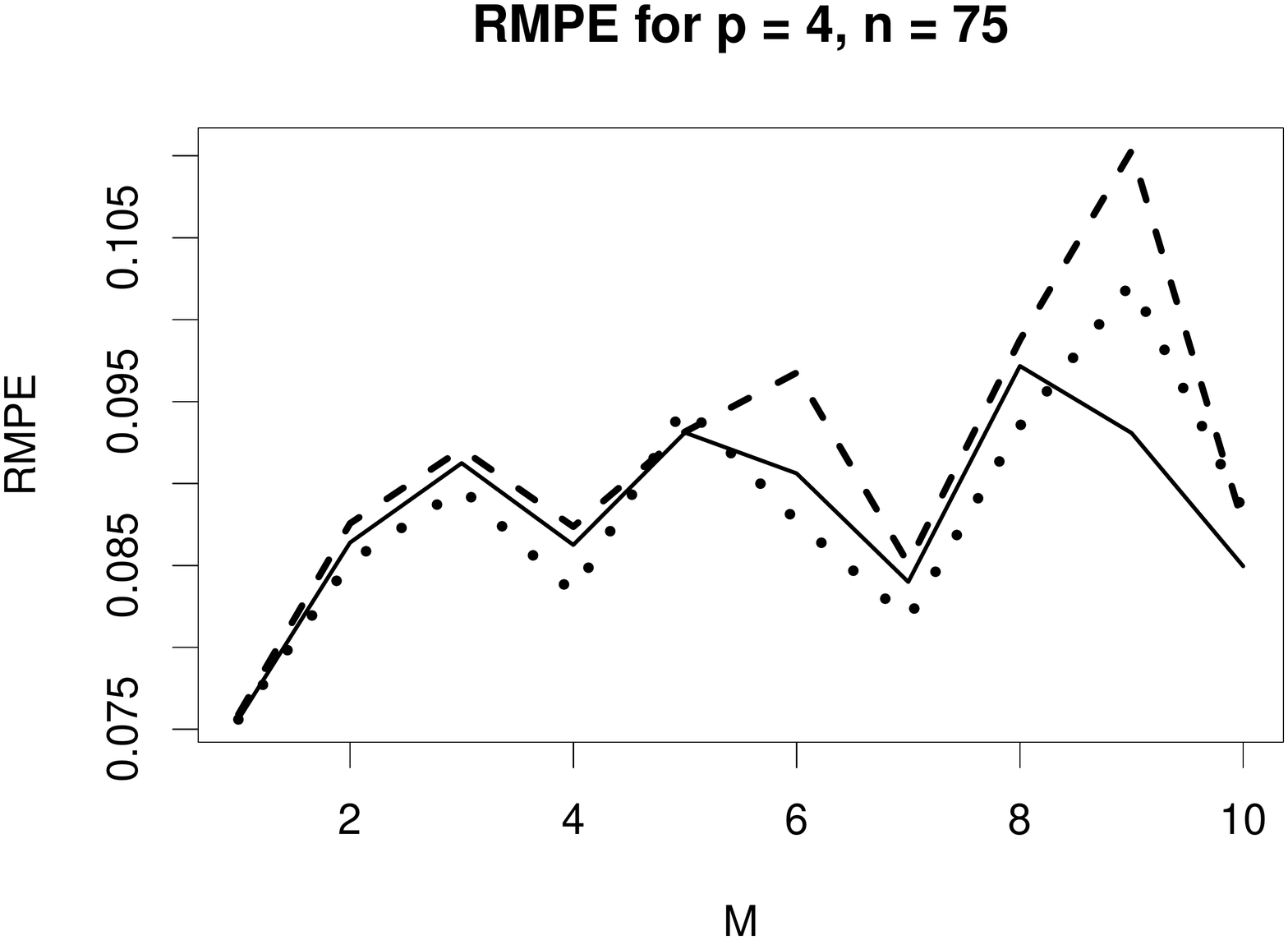}}}
\subfigure[]{\label{fig:rmpeexparb}{%
  \includegraphics[height=0.25\textheight]{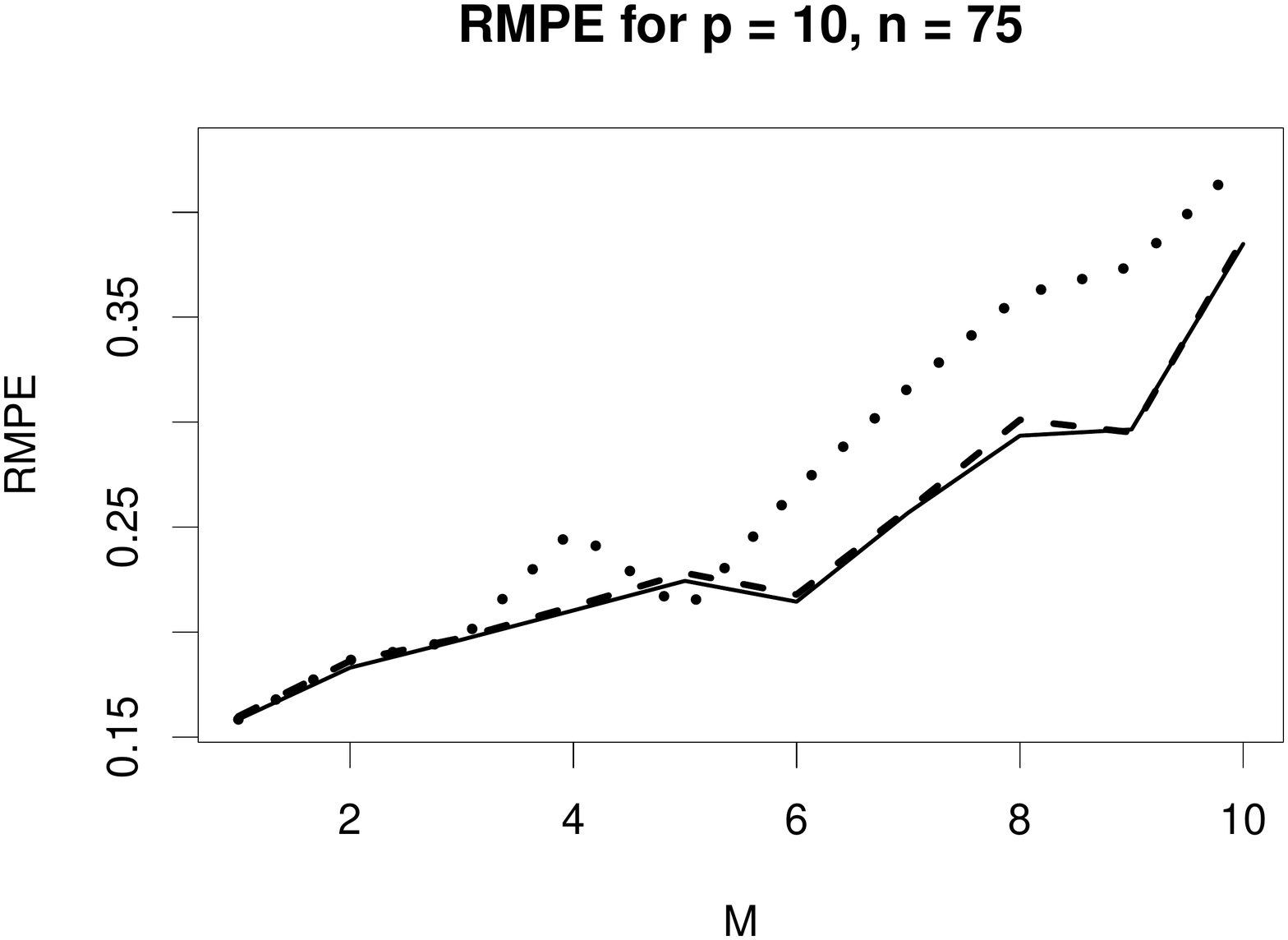}}}
\subfigure[]{\label{fig:rmpeexparc}{%
  \includegraphics[height=0.25\textheight]{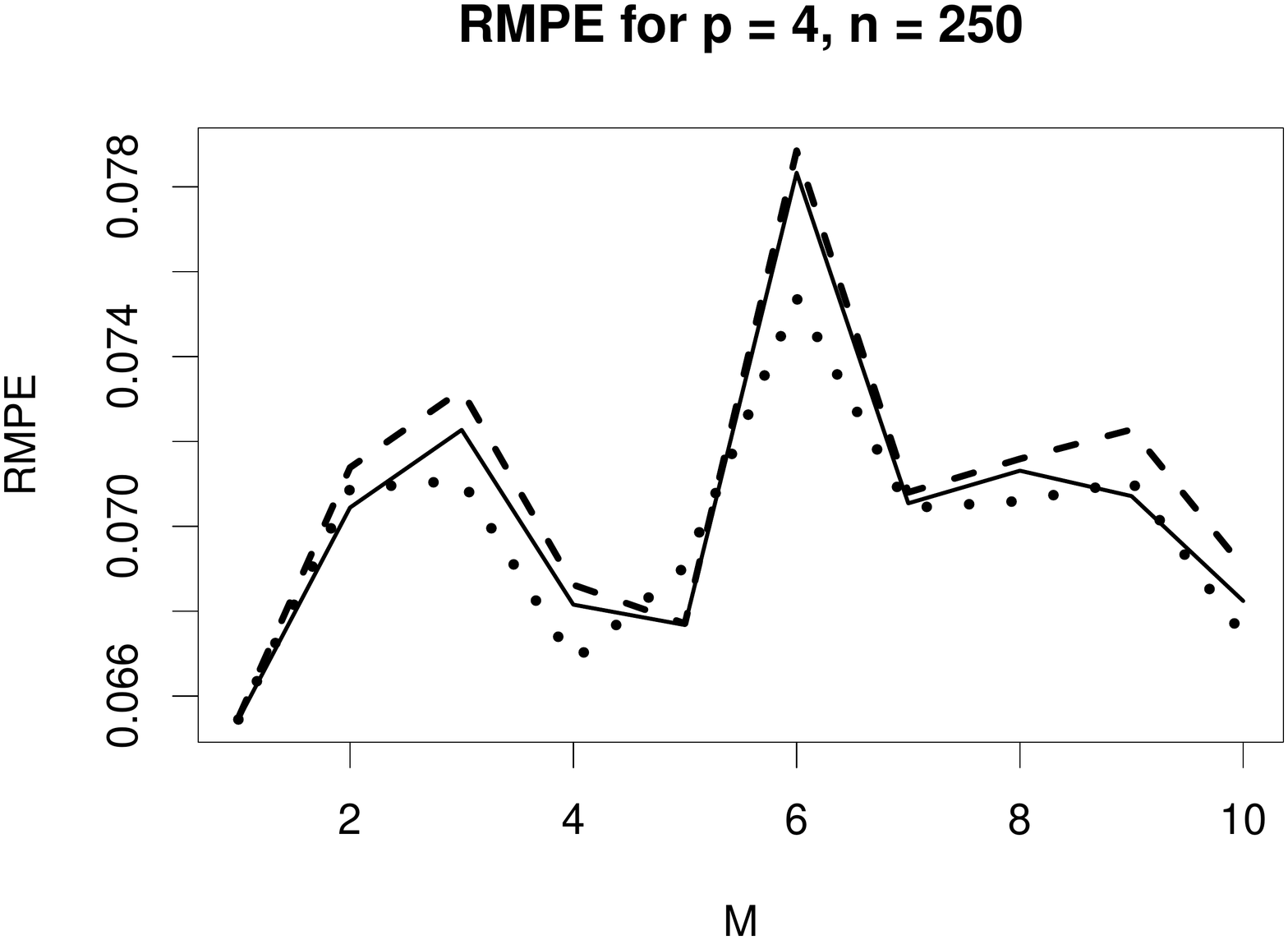}}}
\subfigure[]{\label{fig:rmpeexpard}{%
  \includegraphics[height=0.25\textheight]{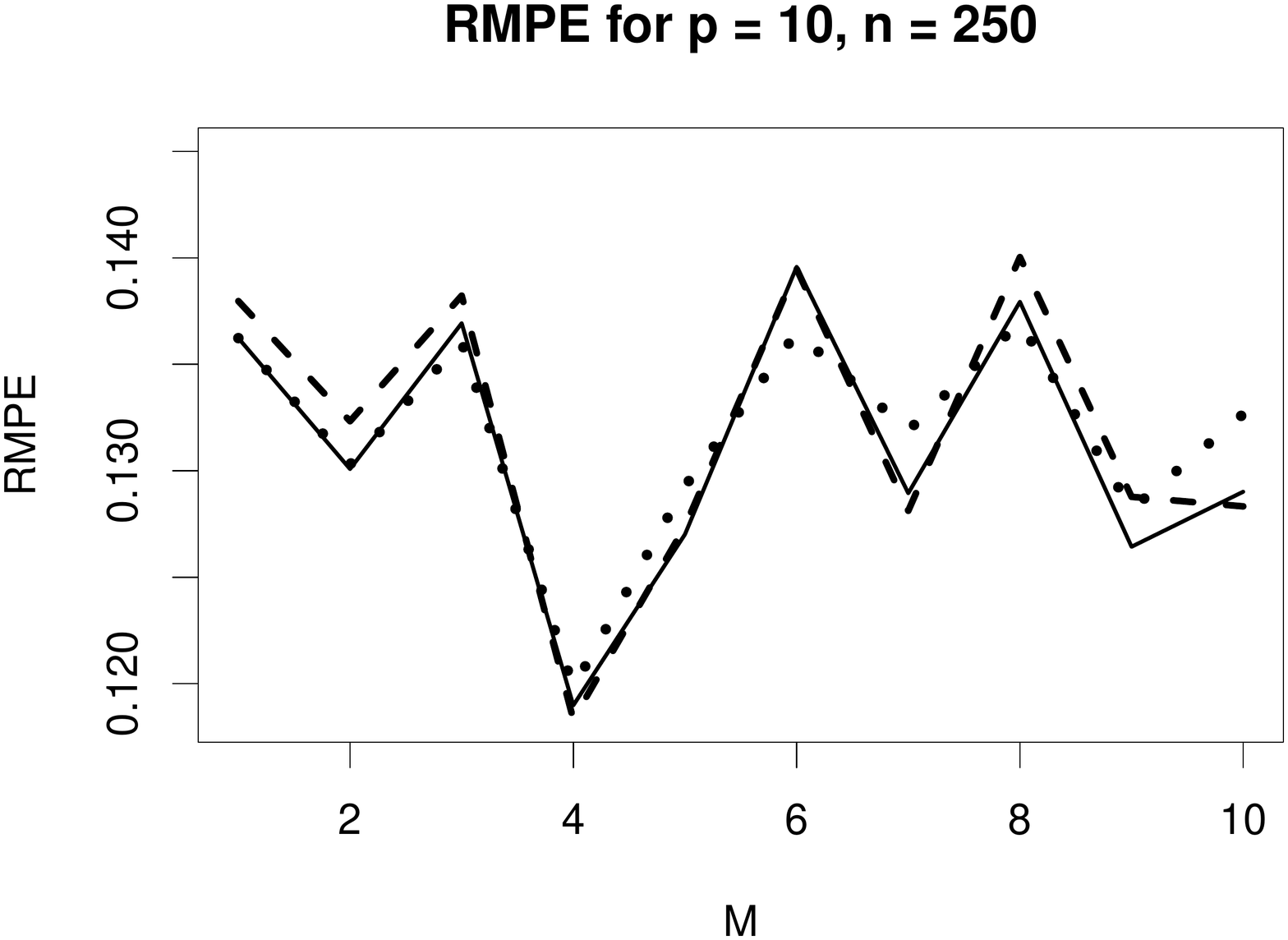}}}
\subfigure[]{\label{fig:rmpeexpare}{%
  \includegraphics[height=0.25\textheight]{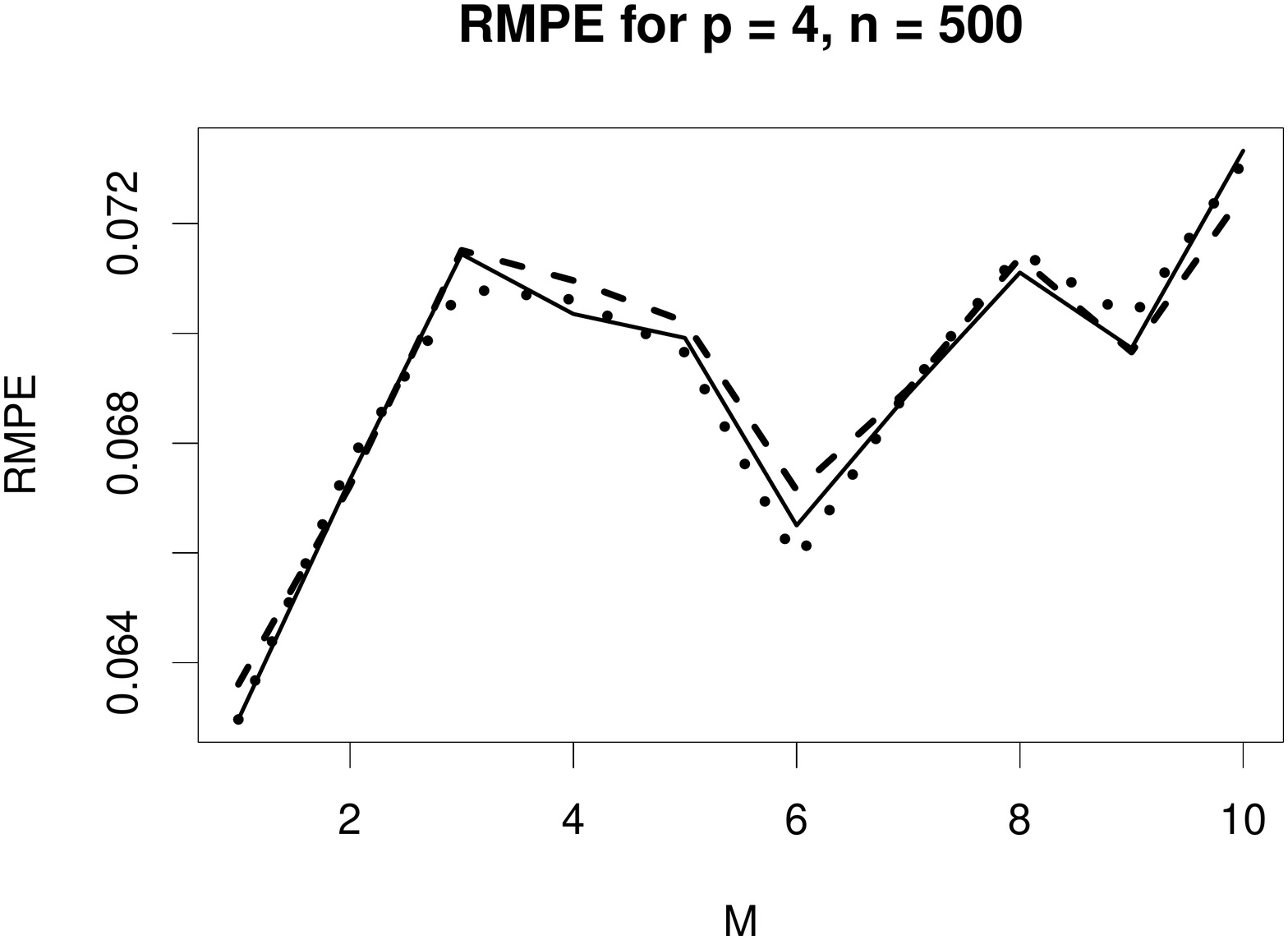}}}
\subfigure[]{\label{fig:rmpeexparf}{%
  \includegraphics[height=0.25\textheight]{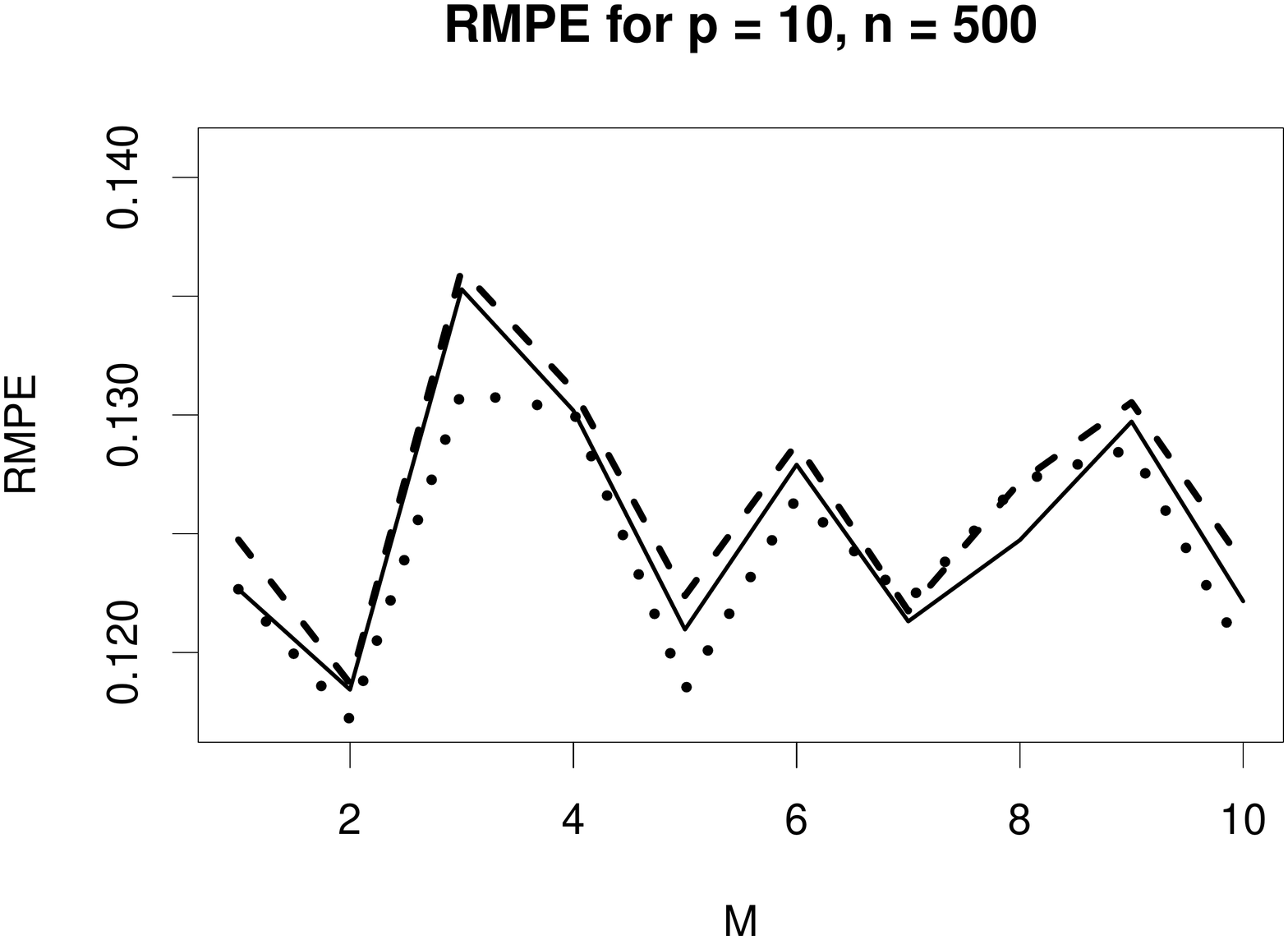}}}
\caption{Plots of the RMPE for Example \ref{ex:expar}.  For each of the plots, the solid line represents the naive forecast, the dotted line represents the bootstrap forecast, and the dashed line represents the multistage forecast. (a) $p=4$, $n=75$, (b) $p=10$, $n=75$, (c) $p=4$, $n=250$, (d) $p=10$, $n=250$, (e) $p=4$, $n=500$, (f) $p=10$, $n=500$.}
\label{fig:rmpeexpar}
\end{center}
\end{figure}
\end{example}

\newpage

\newpage

\newpage

\newpage
\begin{table}[ht]
\caption{Root squared prediction error for $M=1,\ldots,10$ of the naive, multistage, and linear AR forecasts. Values in bold indicate the smallest prediction error for that value of $M$.}
\begin{center}
\scriptsize
\begin{tabular}{|c|cccccccccc|} \hline
  & \multicolumn{10}{|c|}{$M$}\\ \cline{2-11}
           & 1              & 2              & 3              & 4              & 5              & 6              & 7              & 8              & 9                & 10           \\\hline
Naive      & \textbf{0.001} & 0.017          & \textbf{0.013} & \textbf{0.014} & \textbf{0.190} & 0.171          & 0.175          & 0.185          & 0.206          & 0.231        \\
Multistage & \textbf{0.001} & \textbf{0.014} & 0.026          & 0.064          & 0.201          & 0.154          & 0.162          & 0.186          & 0.209          & 0.222        \\
Linear AR  & 0.203          & 0.210          & 0.227          & 0.266          & 0.204          & \textbf{0.111} & \textbf{0.080} & \textbf{0.034} & \textbf{0.006} & \textbf{0.027}\\\hline
\end{tabular}
\end{center}
\label{tab:table1}
\end{table}

\end{document}